%% file: Main.tex
\documentclass[manuscript, nonacm]{acmart}

\usepackage{multirow}
\usepackage{graphicx} 
\usepackage{tabularray}
\usepackage{multicol}
\usepackage{booktabs}
\usepackage{appendix}
\usepackage{caption}
\usepackage{subcaption}
\usepackage{enumitem}
\usepackage{array}
\usepackage{amsmath}
\usepackage{tabularx, booktabs}
\usepackage{tcolorbox}
\usepackage{float}
\usepackage{array}       
\usepackage{rotating}    
\AtBeginDocument{%
  }

\setcopyright{acmlicensed}
\copyrightyear{2018}
\acmYear{2018}
\acmDOI{XXXXXXX.XXXXXXX}
\acmConference[Conference acronym 'XX]{Make sure to enter the correct
  conference title from your rights confirmation email}{June 03--05,
  2018}{Woodstock, NY}
\acmISBN{978-1-4503-XXXX-X/2018/06}

\begin{document}
\input{commands.tex}

\title{CrisisNews: A Dataset Mapping Two Decades of News Articles on Online Problematic Behavior at Scale}

\author{Jeanne Choi}
\authornote{Both authors contributed equally to this research.}
\email{jeannechoi@kaist.ac.kr}
\author{Dongjae Kang}
\authornotemark[1]
\email{jackkang3780@kaist.ac.kr}
\affiliation{%
  \institution{KAIST}
  \city{Daejeon}
  \country{Republic of Korea}
}

\author{Yubin Choi}
\email{yubinchoi@kaist.ac.kr}
\affiliation{%
  \institution{KAIST}
  \city{Daejeon}
  \country{Republic of Korea}
}

\author{Juhoon Lee}
\email{juhoonlee@kaist.ac.kr}
\affiliation{%
  \institution{KAIST}
  \city{Daejeon}
  \country{Republic of Korea}
}

\author{Joseph Seering}
\affiliation{%
  \institution{KAIST}
  \city{Daejeon}
  \country{Republic of Korea}
}
\email{seering@kaist.ac.kr}

\renewcommand{\shortauthors}{Choi, Kang et al.}

\begin{abstract}
    As social media adoption grows globally, online problematic behaviors increasingly escalate into large-scale crises, requiring an evolving set of mitigation strategies. While HCI research often analyzes problematic behaviors with pieces of user-generated content as the unit of analysis, less attention has been given to \textit{event-focused} perspectives that track how discrete events evolve. In this paper, we examine \textit{social media crises}: discrete patterns of problematic behaviors originating and evolving within social media that cause larger-scale harms. Using global news coverage, we present a dataset of 93,250 news articles covering social media-endemic crises from the past 20 years. We analyze a representative subset to classify stakeholder roles, behavior types, and outcomes, uncovering patterns that inform more nuanced classification of social media crises beyond content-based descriptions. By adopting a wider perspective, this research seeks to inform the design of safer platforms, enabling proactive measures to mitigate crises and foster more trustworthy online environments.
\end{abstract}




\received{20 February 2007}
\received[revised]{12 March 2009}
\received[accepted]{5 June 2009}

\maketitle

\input{01_Introduction.tex}
\input{02_Related_Work}

\input{03_Dataset.tex}
\input{04_Analysis.tex}
\input{05_Discussion.tex}
\input{06_Limitation.tex}
\input{07_Conclusion.tex}

\bibliographystyle{ACM-Reference-Format}
\bibliography{sample-base}

\input{08_Appendix.tex}

\end{document}

%% file: commands.tex
\newcommand{\candidate}{\textbf{crisis candidate}}
\newcommand{\initial}{initial dataset}
\newcommand{\smcdata}{CrisisNews}
\newcommand{\anndata}{annotated dataset}
\newcommand{\smc}{}
\newcommand{\stakeholder}{\textsf}
\newcommand{\category}{\textsf}

%% file: 01_Introduction.tex
\section{Introduction}
As the usage of social media continues to grow, so too has the prevalence of problematic online behaviors such as misinformation, hate and harassment, and extremism. Problematic online behaviors take a diverse range of forms, involve many stakeholders, and can lead to a multitude of negative effects, both within and beyond the host platforms. 
To effectively address these risks, it is essential to understand how such behaviors originate and evolve within digital social spaces. Much research in HCI and related fields has studied problematic online behaviors with \textit{content} or \textit{users} as the unit of analysis, and some research has attempted to characterize and contextualize larger incidents, but relatively little research has attempted to conduct comparative analyses of \textit{crises} as a whole.

To understand these events, we draw from the field of crisis informatics to frame the concept of a \smc{\textbf{\textit{social media crisis}}}, which we define as: \textit{an event characterized by online problematic behaviors that originates and evolves primarily within social media and where the structures of social media serve as a catalyst, reaching a level that necessitates larger-scale intervention.} 
This concept of a \smc{social media crisis} emphasizes that severe events in social media emerge and escalate in ways that are uniquely shaped by platforms’ social dynamics and affordances. We do not claim that the study of crises on social media is a novel concept; a substantial body of prior research has examined the kinds of problematic behaviors that happen in online spaces at various scales~\cite{CovidMisinfo01, PrevalenceReddit, Kim2022CSCW}. 
However, we argue here that, as a complement to these types of work, there is value in developing a comparative science of crises on social media where the focus of the analysis is on how different social media crises~---~independent of content type~---~evolve differently over time as different stakeholders and platform structures become involved. This perspective, focusing on crises as the unit of analysis, advances a theoretical understanding of the dynamics of problematic behaviors at scale in digital environments. By foregrounding comparison across crises, our approach highlights patterns that might otherwise remain obscured, enabling both researchers and practitioners to build more effective responses. This approach parallels (or could be seen as extending) traditional crisis informatics, which has emphasized the role of social media in observing, interpreting, and responding to primarily offline crises.

In order to support a more systematic, comparative approach to studying \smc{social media crises}, we present \textit{CrisisNews}: a dataset of news articles on crises endemic to social media, providing a focused lens to study how these crises originate, escalate, and ultimately require intervention. Our dataset comprises over 93,250 crisis articles extracted from international news coverage between 2004 and 2023. Through a combination of keyword filtering, GPT-4-assisted labeling, and semantic merging, we identified incidents that both began and intensified on social media platforms. We then conducted detailed annotations on a statistically representative subset of 1,354 events, categorizing each across multiple dimensions based on information provided in the news articles, including stakeholder roles, online problematic behavior, platform involvement, and outcomes.

Through this dataset, we provide a deeper understanding of the nature of crises on social media, focusing on the analysis of patterns across online problematic behaviors or stakeholders in a crisis. While we recognize that a dataset of news articles carries inherent bias, as journalists do not cover a representative sample of crises, we argue that this dataset offers a significantly broader window into the types of crises that unfold on social media than has previously been available. Our findings offer actionable insights for anticipating potential \smc{social media crises} driven by social media users and their behavior, and can inform the design of safer, more trustworthy social media environments. Such an outcome is possible only by aggregating many crises across different content types and comparing their evolution, which we hope to facilitate with this dataset.

In this paper, we make the following contributions: \vspace{-0.5em}
\begin{itemize}
\item We introduce an operational definition of a \smc{social media crisis} for bounding analysis, grounded in publicly recognized incidents that have demonstrable impact not only to users of social media but also to their broader communities or societies.
\item We build a structured, 20-year longitudinal dataset of \smc{social media crises}, annotated across multiple dimensions—including stakeholder roles, behavior types, platforms, and outcomes—that supports multidimensional analysis of crisis patterns and long-term trends.
\item We utilize a scalable and generalizable pipeline for constructing a macro-level dataset of \smc{social media crises} based on global news sources, enabling systematic discovery beyond keyword-based or platform-internal methods.
\end{itemize}

Together, these contributions lay the groundwork for a comparative science of \smc{social media crises}. Our research provides both a theoretical lens and an empirical foundation for designing interventions that mitigate harms and foster healthier digital environments.



%% file: 02_Related_Work.tex
\section{Related Work}
This section reviews key literature under three focal areas: (1) understanding the landscape of social media in crisis informatics, (2) current approaches of analyzing online problematic behaviors, and (3) approaches to analyzing journalism in HCI research.

\subsection{Social Media in Crisis Informatics}
Crisis informatics, an interdisciplinary field that ``includes empirical study as well as socially and behaviorally conscious ICT development and deployment''~\cite[p.~9]{palen2007crisis}, has traditionally focused on disaster events and emergencies~\cite{kreps1984sociological}. These cases are often defined as an occurrence of an unpredicted event that disrupts stakeholder expectations and may cause substantial harm to social stability~\cite{coombs2007}, which affects many people and requires immediate response~\cite{crisis_1972}. Early crisis informatics research centered on offline crises such as natural disasters~\cite{perry2001facing, fire08} and incidents like terrorist attacks~\cite{moving08}, treating social media mainly as a tool to support crisis management~\cite{crisis_socialmedia_spread}. 

With the rise of social media, crisis communication research in HCI has observed how people leverage features of social media to cope with or respond to crises. Topics include how social media users work to mitigate the impact of crisis events, including long-term community efforts~\cite{longterm15} and emotional support to relieve fear during crises~\cite{DonbasMobile, crisisemoji}. Researchers have investigated social media usage across a wide variety of crisis types, ranging from terror attacks~\cite{terror11, terror13} such as wildfires~\cite{AustraliaFire}, hurricanes~\cite{HurricaneHCI, HurricaneHarvey}, and earthquakes~\cite{JapanEarth}, showing how content shared online can improve situational awareness and aid risk communication by experts. Researchers have explored social media’s role in a similar fashion for human-caused crises: for example, a study of Facebook use in Myanmar during periods of civil unrest showed that individuals leveraged both social networks and messaging apps to rapidly share information about protests and social campaigns, facilitating collective action under repressive conditions~\cite{MyanmarWar}. These HCI studies underscore the extensive roles of social media in crisis contexts, from coping and support to collective sense-making.

A large body of more recent work has analyzed users' experiences on social media specifically during the COVID-19 pandemic, where they used social media to share stress and concerns related to the pandemic~\cite{DistressDisclose, RegretCovid}. This literature has also analyzed COVID-related misinformation in depth (e.g.,~\cite{CovidMisinfo01, CovidMisinfo02}). This work showcases the mixed role of social media in crisis mitigation~---~while sharing information on social media can have powerful protective and restorative effects, social media is also a significant vehicle for misinformation and rumor-spreading.

This prior literature in crisis informatics shows how social media is utilized during offline crises~---~ranging from wildfires to terrorist attacks to pandemics~---~discussing how online tools support information flow, coordination, and public engagement around real-world disaster events.
However, relatively little research in crisis informatics has focused on crises \textit{endemic to social media}, despite the clear parallels. We argue that, with the ever-increasing penetration of social media platforms, online-native crises are both inevitable and consequential, and they warrant more attention from the field of crisis informatics, and the dataset we provide in this paper is an attempt to facilitate this expansion.

\subsection{Studying Problematic Behaviors on Social Media}
Online harm is a broad concept, including a wide range of malicious or abusive behaviors in digital spaces ranging from individual attacks such as harassment, hate speech, or cyberbullying~\cite{tspa}, to group-coordinated attacks like hacking and community-wide disinformation campaigns~\cite{CovidMisinfo01, disinfocscw, RussiaMisinfo}. A growing body of research reveals the serious psychological and social consequences of problematic online behaviors. Victims of hate speech and cyberbullying report depression, anxiety, and in some cases, suicidal ideation~\cite{DepressionHate, SelfharmBully01, SelfharmBully02}. Moreover, the persistence and amplification of harmful content on social media re-traumatizes users, even long after the initial incident~\cite{Trauma}. Beyond emotional injury, there can be economic consequences~\cite{steiger2021psychological}; for example, creators facing hate may lose sponsorships due to reputational risk~\cite{AbleistHate, creator22}. Although such behaviors are typically addressed in platform policies and community content moderation guidelines due to these consequences, their scale, diversity, and evolution make it difficult to proactively predict their emergence and evolution~\cite{modscale}.


A wide variety of research in HCI and CSCW has focused on understanding different types of problematic behaviors, but this research largely focuses on problematic behaviors with either content or users as the unit of analysis, with a few exceptions that evaluate behaviors at a larger scale, e.g., measuring prevalence across a platform~\cite{PrevalenceReddit}. Many studies focus on particular categories of content or community contexts for online problematic behaviors.
For example, recent work across the HCI field has examined problematic behaviors at the micro level. 
On Twitter, studies have analyzed conversational structures to forecast toxicity~\cite{Saveski2021WWW}, investigated when online criticism and ``calling out'' become harassment~\cite{Kim2022CSCW}, and evaluated ``soft-moderation'' interventions such as misinformation warning labels and Community Notes to estimate their short-term effects on engagement~\cite{Papakyriakopoulos2022WWW, Kumar2023WWW, Chuai2024CSCW}. 
While these studies offer detailed insights into toxicity, harassment, and misinformation, they remain largely fragmented~---~typically bounded to a single platform, a single behavior, and limited time horizons.

Despite the clear value of these studies, few have focused on a comparative analysis of crises at larger scales. While it is important and necessary to understand hate speech at a granular level, it is also important to understand how coordinated hate campaigns evolve.\footnote{See~\cite{HateRaidsEchoes} for an example of one such crisis-level evaluation.} Problematic behaviors can be understood and compared not only as isolated actions, but also as part of broader, evolving systems of harm. Such a shift would support more proactive and preventative strategies, something our conceptualization of \smc{social media crises} aims to enable. By collecting a dataset that spans 20 years of journalism on \smc{social media crises}, we aim to provide a foundation for the expansion of such a comparative science. 

\subsection{Journalism and HCI Research}
News has long served as both a methodological testbed and an empirical backdrop for HCI and CSCW research. Prior work has drawn on news corpora to examine patterns and perception of bias~\cite{GenderNews, BIASsist, MediaBias}, evaluate critical thinking and argumentation skills~\cite{MarvistaReading, ReviewAid, DataDive}, and investigate behavioral dynamics such as algorithmic personalization~\cite{AgencyPersonalized} and headline engagement~\cite{HeadlineAnnotation}. Journalistic accounts have also provided an ecologically valid stimulus in experiments on behavior and perception, allowing researchers to study how individuals process information in realistic contexts~\cite{NewsImmigrants, CognitiveCOVID}. Beyond experimental use, news content has been analyzed to investigate the framing of different topics. For instance, HCI and CSCW scholars have examined mainstream media portrayals of mental health research and discourse~\cite{HeadlineMental, MoralMental} and AI technologies~\cite{AIFrame}. Such inquiries are motivated by the fact that news provides systematically produced, publicly accessible, and wide-reaching content. The combination of credibility, accessibility, and impact makes news a valuable resource in conducting human-centered research.

While news has often been studied as the subject in its own right, we focus on its potential as an anchor for mapping broader social phenomena. In HCI, researchers have used news coverage to contextualize online discourse and platform dynamics~\cite{LifecycleNews, AlgorithmsNews}. In adjacent fields, news has served as a rich empirical source for studying social structures or significant events, such as political polarization~\cite{QUOTUS}, public health responses~\cite{MiningHealth}, and natural disasters~\cite{DisasterNews}. Its utility for analyzing social phenomena stems from two distinctive qualities: (1) topicality, capturing issues of urgent public concern as they unfold~\cite{MediatingMedia}, and (2) structured narrative form, which provides coherence and comparability across diverse events and outlets~\cite{MythNews, RealityNews, NewsDiscourse}. These qualities have enabled prior research to track unfolding global crises, such as COVID-19, where scholars combined news coverage with online data to examine public sentiment~\cite{NewsSentiment} and the spread of misinformation~\cite{ReCOVery}.

Despite this promise, few studies have explored online problematic behaviors as critical social phenomena, especially through the lens of news coverage. Journalistic reports offer inherent narrative features~---~fundamental interrogatives (who, what, when, where, why, and how), temporal sequencing, and causal explanations~---~that can help chart how crises are constructed and evolve across public discourse. Importantly, news has close ties to the social media space, serving simultaneously as messenger and message: it both informs online discussion and is itself disseminated, debated, and reframed across platforms. We argue that news as a data source has the potential to capture the macro-scale patterns of problematic behaviors on social media. Leveraging global news coverage thus provides a structured and comprehensive empirical basis for investigating the emergence, escalation, and societal impact of \smc{social media crises}, grounding such analyses in a broader socio-technical context. We propose that news sources have underutilized methodological value as a scaffold to building event-centered datasets, supporting the creation of coherent datasets of \smc{social media crises}, enabling comparative and longitudinal analyses of online harms.

%% file: 03_Dataset.tex
\section{Dataset Creation}
In this section, we introduce the \textit{CrisisNews} dataset and outline the steps for (1) defining inclusion criteria, (2) collecting data, (3) filtering the data, and (4) performing sample analysis by manual annotation. Though the primary purpose of this paper is to introduce this dataset~---~not to analyze it in full~---~we demonstrate the types of analysis that might be useful for taking steps toward a comparative science of \smc{social media crises}.\footnote{The dataset can be viewed at \url{https://crisis-news.netlify.app/}}

\subsection{Definition of \smc{Social Media Crisis}}
For the purpose of creating a dataset of news articles related to crises on social media, we must first settle on a definition of such crises. In doing so, we turn to crisis informatics to adopt criteria traditionally used to define a crisis. A crisis may be understood as the occurrence of an unforeseen event that disrupts essential stakeholder expectations and has the potential to cause substantial harm to stability, objectives, and overall performance~\cite{crisis_comb_2007}. 
Per Hermann, a crisis has three defining traits: high threat, limited decision time, and surprise~\cite{crisis_1972}. Due to its sudden high-impact, a crisis poses a severe threat to essential values or goals such as public safety or an organization’s viability, and demands immediate response under conditions of deep uncertainty~\cite{Boin2017, crisis_comb_2007}. The defining characteristics of crisis, including a significant threat, a sense of urgency, and pervasive uncertainty, underscore the demand for timely communication and information exchange strategies at each stage of the crisis’s life cycle~\cite{ulmer2022crisis, Boin2017}. 

Based on this definition and characteristics of crisis, we attempt to define social media-endemic crises as \smc{\textit{social media crises}}. First, we identify social media as web-based applications and interactive platforms that facilitate the creation, discussion, modification, and exchange of user-generated content~\cite{socialmediadef}. Note that we do not limit our definition of social media to social networks like Facebook but also include various other platforms that enable diverse forms of digital interaction and content creation~\cite{socialmediadef2} such as blogs or messaging apps. In turn, a \smc{social media crisis} is then defined as an event characterized by online problematic behavior that originates and evolves primarily within social media and where the structures of social media serve as a catalyst, reaching a level that necessitates larger-scale intervention. Though \smc{social media crises} may be triggered both by offline and/or online events, the impact is mainly situated on social media. This definition emphasizes that while crises may arise from varied triggers, their evolution and impact within social media are uniquely shaped by platform dynamics, making them analytically distinct from traditional crises. 



\subsection{Data Collection}

\begin{figure}
    \centering
    \includegraphics[width=\linewidth]{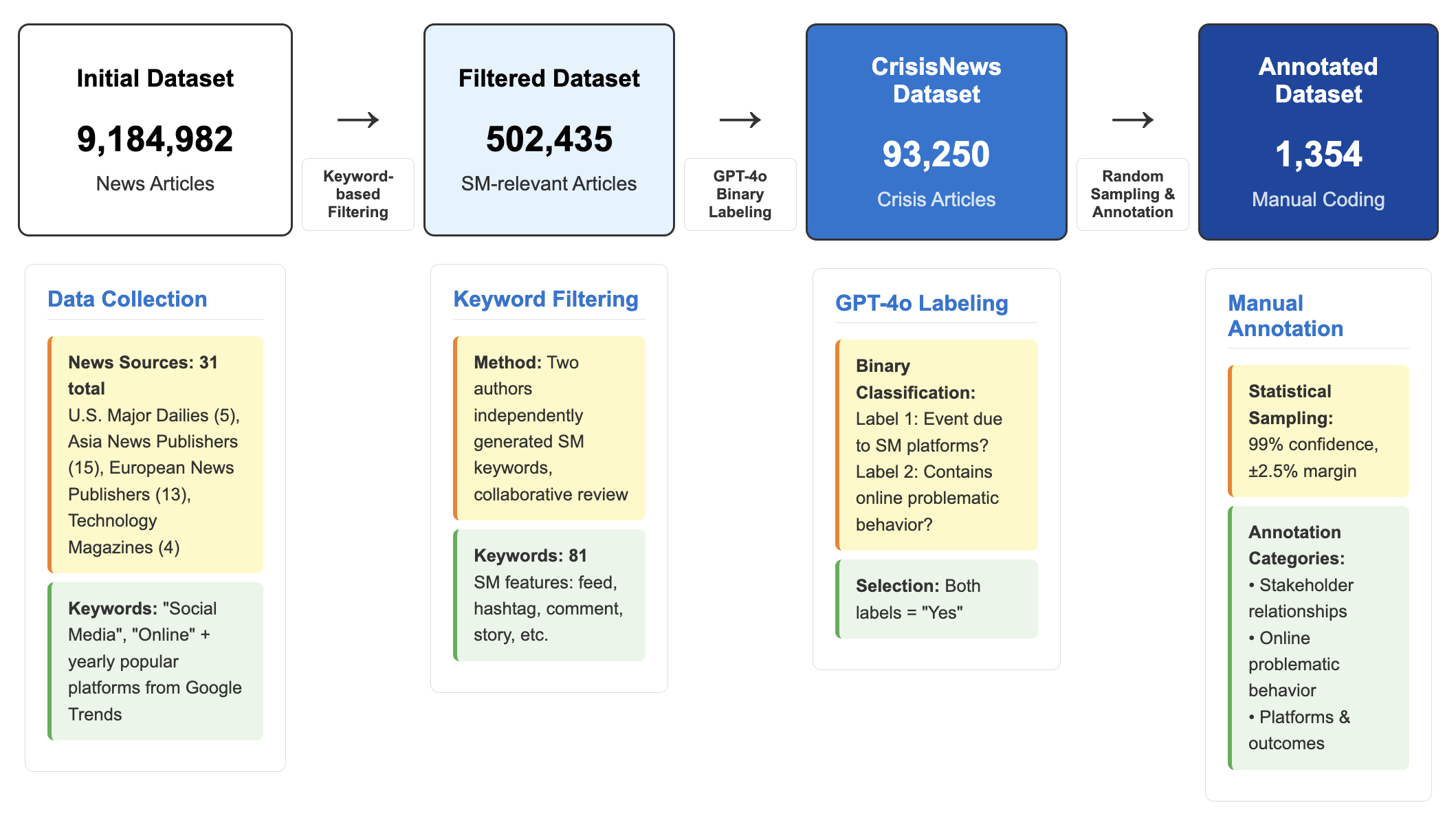}
    \caption{CrisisNews Dataset Creation Pipeline}
    \Description{The image shows the dataset creation pipeline for CrisisNews dataset. The initial dataset on the left shows 9,184,982 news articles, with the data collection method below, including 31 news sources and keywords such as "Social Media" and "Crisis". The next column shows the keyword-based filtered dataset at 502,435 articles using independently generated 81 social media keywords. The next column shows the final CrisisNews dataset at 93,250 crisis articles using GPT-4o labeling, and the final column shows the annotated dataset of 1354 articles using stakeholder relationships, online problematic behavior, and platforms and the outcomes as the annotation categories.}
    \label{fig:placeholder}
\end{figure}

We collected data on \smc{social media crises} from news sources, as news both reflects broader societal phenomena~\cite{QUOTUS, MiningHealth, DisasterNews} and maintains a close interrelationship with events that unfold in social media spaces~\cite{NewsSentiment, ReCOVery}. Candidate news articles for the dataset from 2004 to 2023 were provided through TDM studio~\cite{ProquestTDMStudio}, a data mining platform that enables researchers to analyze large-scale scholarly content. Here, we refer to the initial unfiltered collection of news articles as the \textit{\initial{}}. While we recognize that there is no single clear beginning for social media crises, we choose 2004 as the starting year primarily due to the growing popularity of what had become termed ``Web 2.0''~\cite{o2005web} and the launch of what might be considered modern social media (e.g., Facebook)~\cite{web20}.

The \initial{} was collected from selected representative news sources shown in Appendix~\ref{app:newspublisher}. The group of US Major Dailies (as defined in TDM studio) was chosen based on its credibility in the United States, and representative news sources from Asia and Europe were also chosen based on overall popularity throughout this time window. In order to supplement this collection, we also included a small set of technology-focused periodicals in order to capture news on more niche computing topics that might include early reporting on social media.

From among the selected news sources, we initially selected articles based on keywords as an input for data retrieval in TDM studio, including ``Social Media'', ``Online'', and also the names of social media platforms that were most popular for each year. The social media platforms were selected from Google Trends,\footnote{https://trends.google.com/trends/} which provides the top search themes of each year. In Google Trends, various categories of search topics (e.g., ``Arts and Entertainment'', ``Health'') are ranked, and to identify popular platforms, we used the category ``Online Community''. Within the category, Google provides both top ``Search topics'' and ``Search queries''; we used ``Search topics'' for constructing the dataset, as it contains the exact names of the social media platforms. Topics in the ``Online Community'' category were ranked based on ``Top'', topics, which contains the five most-searched items. We used social media platforms that were listed in the ``Top'' topics for each year, except for the years from 2004 to 2006. For this time period, there were fewer than five platforms listed in ``Top'', so we supplemented the list with platforms appearing in the ``Rising'' topics. The social media platforms that were used as keywords are shown in Appendix~\ref{tab:platform keyword}. The total number of news articles in the \initial{} after performing keyword filtering was 9,184,982.

\subsection{Data Filtering}

From the \initial{}, we conducted the three steps below to filter the data into the final dataset.

\subsubsection{Filtering by social media-related keywords}
Although our initial selection targeted articles related to major social media platforms, subsequent review of the dataset revealed the presence of a significant number of irrelevant articles. Thus, to identify news articles directly relevant to social media among the articles in \initial{}, we employed a keyword-based filtering strategy. Each of the two first authors independently generated a comprehensive list of keywords associated with social media. These keywords encompassed the core features and interaction modalities in the platforms (e.g., feed, story, hashtag, comment) in order to capture articles that only mention specific features in a platform instead of the name of the platform. Following this, the two authors collaboratively reviewed the keyword lists to extract the overlapping entries and further refined the selection through discussion. We have included the list of social media platforms used for the initial selection of data for consistency. This process yielded a final set of 172 keywords, which were used to filter the collected dataset and isolate articles with strong relevance to social media. The complete list of keywords is shown in Appendix~\ref{tab:keyword}. In this step, we also removed articles with titles of fewer than three words and exact duplicate articles. This resulted in a dataset of 502,435 articles.

\subsubsection{Labeling the articles}
After applying keyword filters to identify relevant content from the \initial{}, each remaining article received binary labels indicating whether (1) the event occurred due to the existence of social media platforms, and (2) the article contains evidence of online problematic behavior. These labels served to refine the dataset into a more focused subset of articles that captured events both enabled by social media infrastructure and involving behaviors that could escalate into societal-level interventions.


We focused our labeling analysis exclusively on article titles rather than full text for several methodological reasons. First, news headlines are specifically designed to encapsulate the core essence of events, serving as concentrated summaries that capture the most newsworthy aspects of incidents. Additionally, our preliminary exploratory review indicated that headlines provide a sufficient signal for our binary classification task, demonstrating that title-based analysis captures the events most relevant to our research objectives while maintaining scalability across our large dataset.

To ensure scalability and consistency in the annotation process, we used the GPT-4o API, a large language model, to generate the labels for each article~\cite{wang2024human,
zhu2023can}. The complete prompt used for this annotation is provided in Appendix~\ref{appendix:prompt}. 
To validate this approach, we compared the GPT labels with the first and second authors' labels for the same task on 100 articles for both binary classification tasks based on the two criteria. The average Cohen's Kappa scores for inter-rater agreement were 0.79 for criteria 1 and 0.75 for criteria 2, indicating substantial agreement. Following this validation, we applied GPT-4o to label the complete dataset. Based on the labeled results, we have selected the articles that were labeled as ``yes'' for both criteria, resulting in the final set of 93,250 articles, which we refer to as the \textit{CrisisNews} dataset.





\subsection{Analysis Method}
For the analysis of the articles, we annotated each event based on major factors of a \smc{social media crisis}.
To perform the analysis, we randomly selected and annotated a sampling of 1,354 articles using simple random sampling based on stakeholder relationships (N=25) and Cochran's formula for large populations~\cite{cochran1977sampling}. This sample size was statistically determined based on a preliminary analysis of 100 articles, which revealed that the largest category of stakeholder relationships represented 15\% of cases. Using a 99\% confidence level with a $\pm$2.5\% margin of error, this sample size ensures robust statistical representation of the full dataset, enabling generalization of our findings across the complete collection of articles.
We refer to the annotated cases as the \textit{\anndata{}}. To ensure research transparency and to facilitate replication, we include direct links to publisher websites in our publicly released annotated dataset.

\subsubsection{Annotation Preparation}
The annotation process was carried out by the two first authors by systematically reading and examining the news articles. Prior to starting the annotation process, we first assessed whether each article qualified as a \smc{social media crisis}, anticipating the presence of error cases within the dataset. If a case was discovered to not be a \smc{social media crisis}, we labeled them among the two categories: (1) \textbf{Non-Crisis}---cases that do not contain any attributes a of \smc{social media crisis}---and (2) \textbf{Relevant to Crisis}---cases that have relevance to a \smc{social media crisis} but do not show clear, explicit online problematic behavior or did not happen in a social online space. Non-Crisis articles usually included error cases that were filtered with keywords that have duplicate meanings. 
While we removed Non-Crisis articles, we chose to include the Relevant to Crisis category in our dataset, as these show how specific affordances or interactions in social media such as anonymity, algorithmic amplification, or online relationships that could potentially lead to \smc{social media crises} as they often escalate into offline consequences~\cite{anony18, anony18_2, anony20}. By including these contexts, \smcdata{} provides a more holistic view of crisis development, foregrounding the socio-technical dynamics that connect digital interactions with real-world risks and outcomes.

The filtering resulted in 1,112 cases of Social Media Crises, 115 cases of Relevant to Crisis, and 127 cases of Non-Crisis. This shows an error rate of 9.38\% for Non-Crisis events. We performed annotation on the remaining 1,227 cases.

\subsubsection{Annotation Process}
For the construction of annotation categories for a \smc{social media crisis}, we performed several rounds of open discussion between the first authors after examining the sample set, discussing the important features of \smc{social media crises} and considering what information is typically obtainable from news articles. After a thorough discussion, the authors derived the annotation categories: (1) Stakeholders involved in the \smc{social media crisis} and their relationship, (2) Type of online problematic behavior; (3) Platform on which the \smc{social media crisis} occurred; and (4) Aftermath of the \smc{social media crisis}.

The authors first asynchronously annotated an identical set of ten events across the four annotation categories. Then the authors engaged in a detailed discussion to set the desired granularity of annotation and to establish common criteria for interpreting specific behaviors and stakeholder groups described in the articles. After reaching a consensus, the annotation protocol was created, and the authors proceeded with the annotation for the rest of the events accordingly. In the process, there were events that did not fit in any existing subcategories. If so, the first authors discussed whether a new subcategory should be added, creating a subcategory when both authors agreed. We explain each category and its subcategories in detail below. The full list of subcategories for each category is in Appendix~\ref{fullsubcat}. 

\begin{itemize}
\item \textbf{Online Problematic Behavior}: For online problematic behavior categorization, we adopted the Abuse Types (AT) framework from the Trust and Safety Professional Association~\cite{tspa} as our foundation, as it provides a standardized framework for defining online problematic behaviors. AT categorizes various online misbehavior into seven large categories, each having several subcategories (total number is written within parentheses): \category{Violent and Criminal Behavior} (5), \category{Regulated Goods and Services} (3), \category{Offensive and Objectionable Content} (3), \category{User Safety} (3), \category{Scaled Abuse} (3), \category{Deceptive and Fraudulent Behavior} (5), and \category{Community-Specific Rules} (2), resulting in a total of 24 subcategories. During the annotation process, we added six subcategories for events that could not be categorized within the existing list~(Table~\ref{addopb}). We thus define a total of 28 subcategories under the seven categories of AT. We exclude the subcategory of \category{Violent \& Criminal Behavior~--~Human Exploitation} and \category{Deceptive and Fraudulent Behavior \& Cybersecurity} as \anndata{} did not contain any relevant cases.

\begin{table}[htbp]
\centering
\small
\caption{A List of Additional Online Problematic Behavior Subcategories Used for Annotation.}
\begin{tabularx}{\textwidth}{@{} l l >{\raggedright\arraybackslash}X @{}}
\toprule
\textbf{Top Category} & \textbf{Subcategory} & \textbf{Explanation} \\
\midrule
Scaled Abuse & Hacking & Events that involve large-scale hacking activities affecting multiple systems, platforms, or user groups. \\
\midrule
Violent and Criminal Behavior & Illegal Behavior & Events that involve criminal behaviors that are illegal and exposed through social media. \\
\midrule
User Safety & Safety Risk of Social Media Overuse & Events that occur due to addiction or overuse of social media. This starts from individual harms that users encounter, such as amplification of negative body image or addiction to social media. \\
\midrule
User Safety & Censorship and Retribution & Events related to censorship from an influential entity on specific content uploaded in social media or even a ban on a specific social media platform, typically by the government. \\
\midrule
User Safety & Personal Information & Events where personal information is threatened or leaked from a social media platform, both intentional or unintentional. \\
\midrule
User Safety & Broken Harmony* & Events where the `harmony' of the society within the social media user group is disturbed due to a sudden surge in social controversy that is unconstructive or even violent.\\
\bottomrule
\multicolumn{2}{c}{*This category is explained in significantly more detail in Section 4.}\\ \\
\end{tabularx}
\Description{The table displays a list of the additionally identified online problematic behavior subcategories used for annotation. On the far left is the top category that the subcategory belongs to, in the middle is the added subcategory, and on the right is the explanation of the subcategory. The subcategories added are Hacking in Scaled Abuse, Illegal Behavior in Violent and Criminal Behavior, Safety Risk of Social Media Overuse, Censorship and Retribution, Personal Information, and Broken Harmony all in User Safety.}
\label{addopb}
\end{table}

\item \textbf{Stakeholders and Stakeholder Relationships}:
We define stakeholders of a \smc{social media crisis} as those who are directly related to the online problematic behavior mentioned in the event. These stakeholders are divided along two criteria: (1) size and (2) impact. For size, the stakeholders are divided by \stakeholder{Individual}, \stakeholder{Group}, and \stakeholder{Social Media Platform}. For impact, the \stakeholder{Individual} and \stakeholder{Group} stakeholders are further divided into \stakeholder{Influential}, which are users who have a large impact on other users (e.g., influencers or politicians), and \stakeholder{Regular}, consisting of the majority of users without extensive reach or influence. Here, size and impact create combinations to illustrate the relative scale of stakeholders and the extent of their influence within the platform, except for \stakeholder{Social Media Platform} as it is not definable along the two categories. In sum, we define five stakeholders: \stakeholder{Influential\_Individual}, \stakeholder{Influential\_Group}, \stakeholder{Regular\_Individual}, \stakeholder{Regular\_Group}, \stakeholder{Social Media Platform}.

The five stakeholders can each be a \stakeholder{Giver} or a \stakeholder{Receiver} of online problematic behavior, combining into a stakeholder relationship. Here, \stakeholder{Giver} is the stakeholder who conducted the online problematic behavior, and \stakeholder{Receiver} is the stakeholder who received harm due to the online problematic behavior. We express the stakeholder relationship using an arrow with a direction ($\rightarrow$), pointing from \stakeholder{Giver} to \stakeholder{Receiver}. For example, if the \stakeholder{Giver} is \stakeholder{Influential\_Individual} and the \stakeholder{Receiver} is \stakeholder{Regular\_Group}, the stakeholder relationship will be expressed as (\stakeholder{Influential\_Individual} $\rightarrow$ \stakeholder{Regular\_Group}).\\

\item \textbf{Platform and Aftermath}: We annotated the social media platform on which the event occurred, as mentioned in the article. However, articles often did not explicitly convey the exact social media platform of the event but rather referred to the platform as ``social media'' or ``SNS.'' We denote such cases as simply ``Social Media''. We found a total of 96 platform categories. The aftermath of the event was annotated only when there was an explicit aftermath or result of the online problematic behaviors explained in the article. We found a total of 31 categories for aftermath.    
\end{itemize}




 

%% file: 04_Analysis.tex
\section{Analysis of the \smcdata{} Dataset}
Based on the annotation, we analyze \smc{social media crises} to examine the common patterns and features found throughout \anndata{}. We first provide an overall statistical analysis on the dataset, followed by unique findings on the characteristics of various stakeholders and their relationships in the \smc{social media crisis}, with a focus on social media platforms as a stakeholder and user influence as the basis factor. Finally, we provide a detailed view of the new subcategory of online problematic behaviors our annotation examined, \category{User Safety~--~Broken Harmony}. Note that in this paper we do not discuss the patterns identified across every labeled category in the annotated dataset. Instead, we highlight a subset of the most interesting trends. This analysis is intended to serve as an example of how this dataset might be used in the future.

\subsection{Overall Statistical Analysis}

In this subsection, we discuss overall statistical analyses of the dataset, describing overall trends in articles per year, articles by publisher, distribution across stakeholder groups, and representation of different problematic behaviors. Note that we do not make any statistical claims about online problematic behaviors \textit{as a whole} as a result of this data. Due to the aforementioned biases, we cannot claim that the dataset covers a fully representative set of all social media crises. Nevertheless, we present this analysis to characterize the overall composition of this dataset.
\subsubsection{Articles per Year}
\begin{figure}
    \centering
    \includegraphics[width=0.7\linewidth]{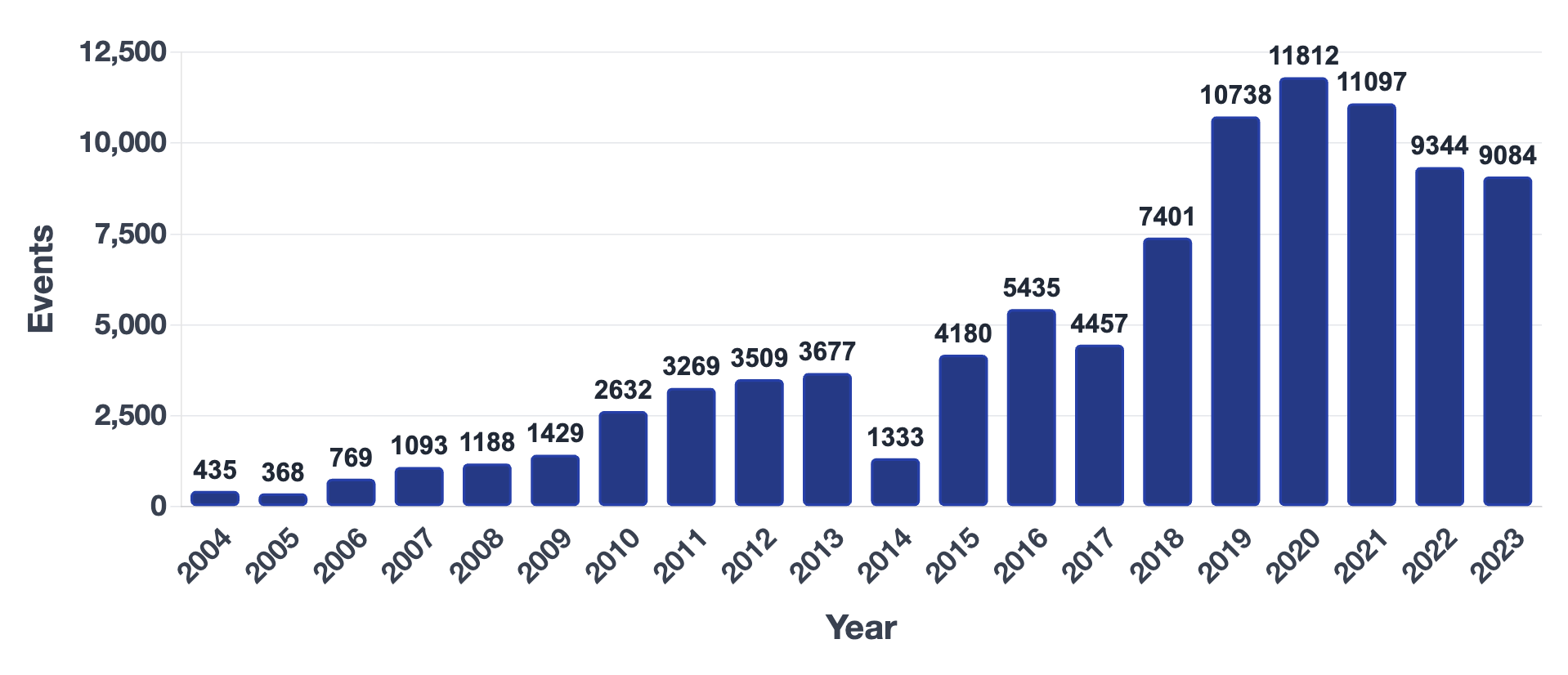}
    \caption{Yearly Distribution of News Articles in CrisisNews}
    \label{fig:enter-label}
    \Description{The figure shows a bar graph over time of the yearly distribution of news articles in the CrisisNews dataset. The x-axis begins from 2004 to 2023. The trends show an upward trend, beginning with 8 in 2004 and peaking at 139 in 2019.}
\end{figure}
In this dataset, we collect news articles that span 20 years from 2004 to 2023. Figure~\ref{fig:enter-label} illustrates the annual frequency of events recorded in our dataset. Overall, there is a marked upward trajectory in the volume of documented events over the two-decade period. The number of events remained relatively low and stable between 2004 and 2009, with annual counts ranging from 368 to 1429. Beginning in 2010, however, the dataset exhibits a sustained increase, having over 2,000 events per year and reaching over 4,000 events annually by 2015. The most substantial growth occurred between 2017 and 2020, showing a peak of 11,812 events in 2020, and the annual count consistently holds over 9,000 in the most recent years. This trend likely reflects both the increase in the use of social media platforms and the growing societal attention to online problematic behaviors over time.

\subsubsection{Articles per News Publisher}
\begin{figure}[h]
    \centering
    \begin{minipage}[b]{0.47\textwidth}
        \centering
        \includegraphics[width=\textwidth]{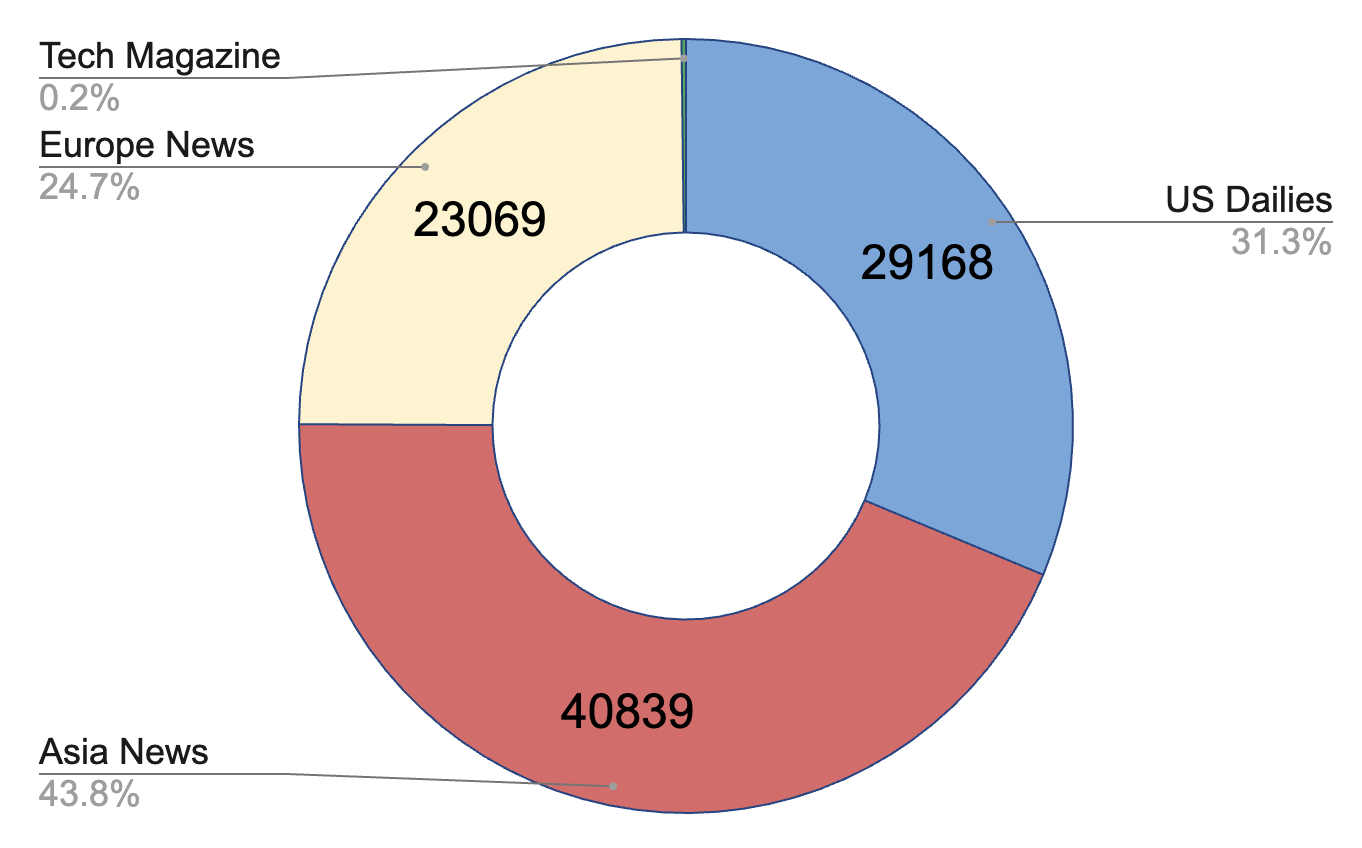}
        \caption{Distribution of Articles by News Publisher Category}
        \Description{The figure shows a pie chart of the distribution of articles by news publisher category, showing the distribution of US Dailies, Asia News, Europe News, and Tech Magazines. Asia News takes up 43.8 percent with 40839 articles, US Dailies take up 31.3 percent with 29168 articles, and Europe News takes up 24.7 percent with 23079 articles. Tech Magazines make up 0.2 percent of the distribution.}
        \label{fig:publisher-continent}
    \end{minipage}
    \hspace{0.05\textwidth} 
    \begin{minipage}[b]{0.47\textwidth}
        \centering
        \includegraphics[width=\textwidth]{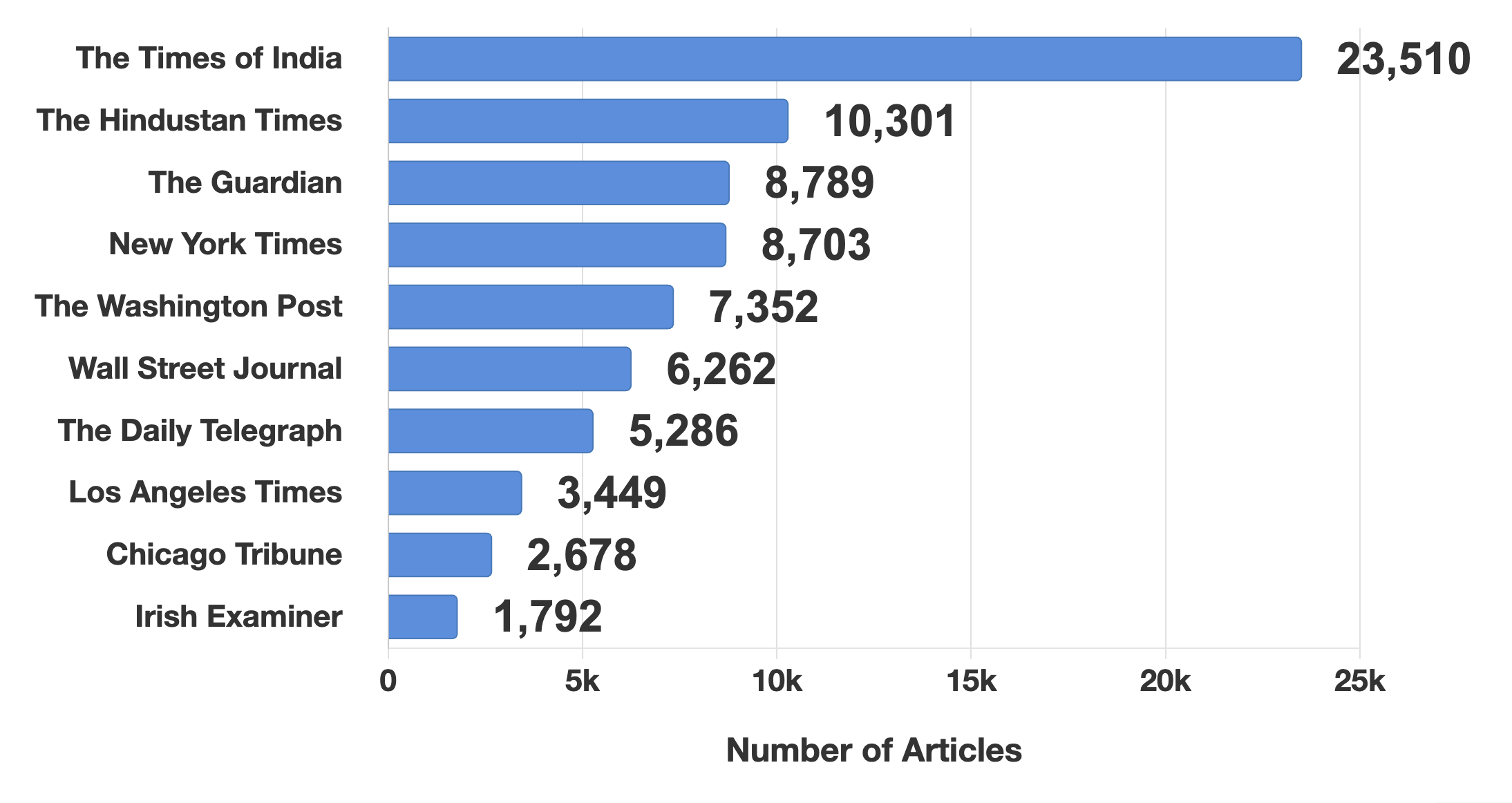}
        \caption{Distribution of Articles by News Publisher (Top 10)}
        \Description{The figure shows a lateral bar graph of the distribution of articles by the top ten most frequent news publishers. The order is Times of India, Hindustan Times, The Guardian, New York Times, The Washington Post, Wall Street Journal, The Daily Telegraph, Los Angeles Times, Chicago Tribune, and Irish Examiner. The Times of India shows much greater number than the other sources.}
        \label{fig:publisher}
    \end{minipage}
\end{figure}
Figures~\ref{fig:publisher-continent} and ~\ref{fig:publisher} summarize the distribution of news publishers represented in the dataset. Figure~\ref{fig:publisher-continent} shows the articles in each regional category, indicating that Asia-based news publishers collectively constitute the largest part of the dataset (43.8\%). US major dailies comprise 31.3\% of all records, while European news publishers account for approximately one quarter (24.7\%). The result from Figure~\ref{fig:publisher-continent} can be explained by Figure~\ref{fig:publisher}, which shows the frequency of events by individual news outlet, highlighting several prominent contributors. The Hindustan Times and Times of India emerge as the two most frequent sources, contributing to the dominance of Asia-based news publishers in the dataset.

\subsubsection{Distribution across Stakeholder Groups}
\begin{figure}[h]
    \centering
    \includegraphics[width=0.84\textwidth]{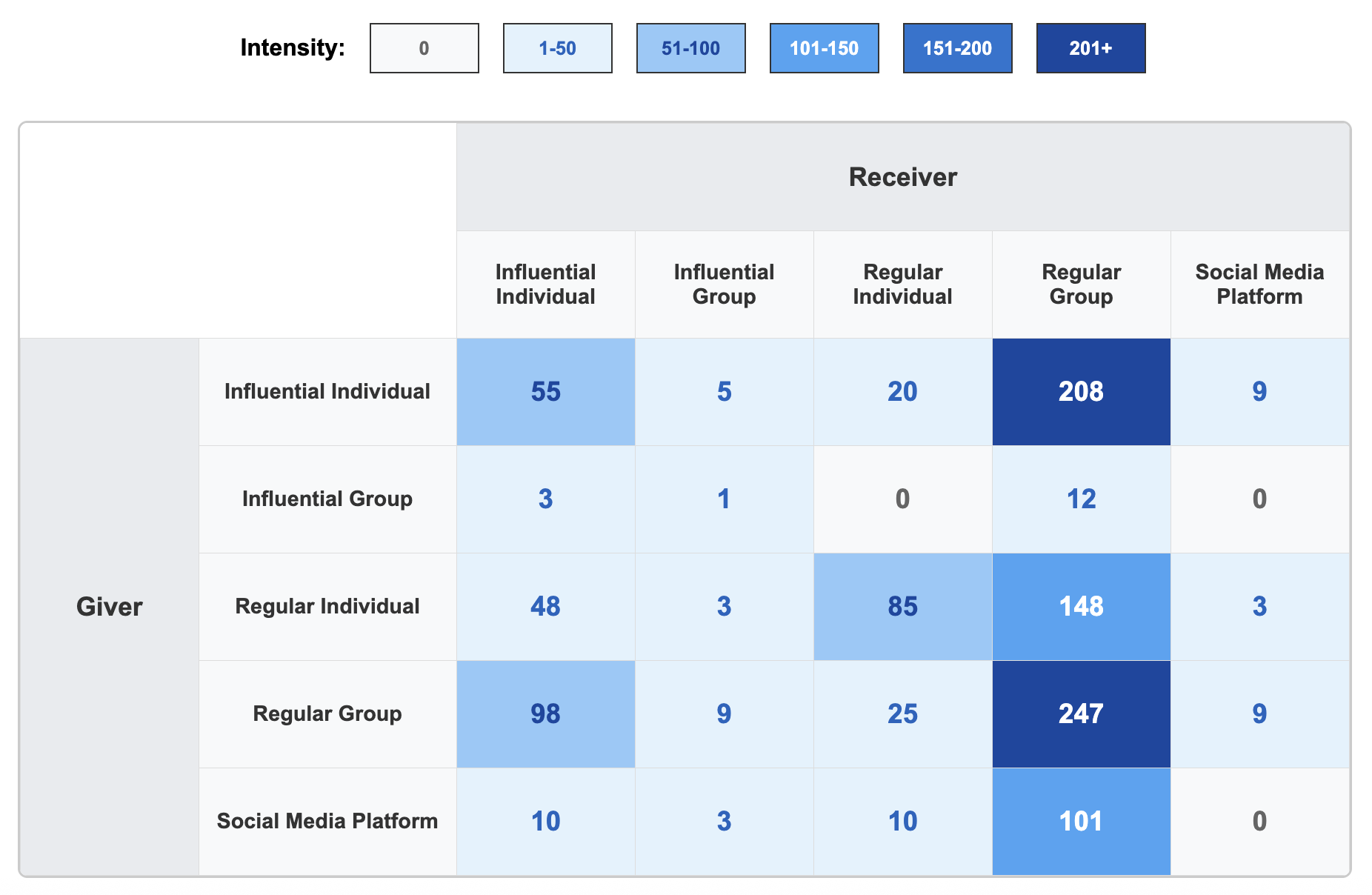}
    \caption{Stakeholder Relationship Matrix (Count)}
    \Description{The figure shows the stakeholder relationship matrix according to the count among the five possible stakeholders of Influential Individual, Influential Group, Regular Individual, Regular Group, and Social Media Platform. The right column is denoted as the Giver, while the top row is denoted as the Receiver. Regular Group to Regular Group is the most frequent at 208 counts, and Influential Individual to Regular Group is second at 208 counts.}
    \label{fig:stakeholder-analysis-count}
\end{figure}
\begin{figure}[h]
    \centering
    \includegraphics[width=0.84\textwidth]{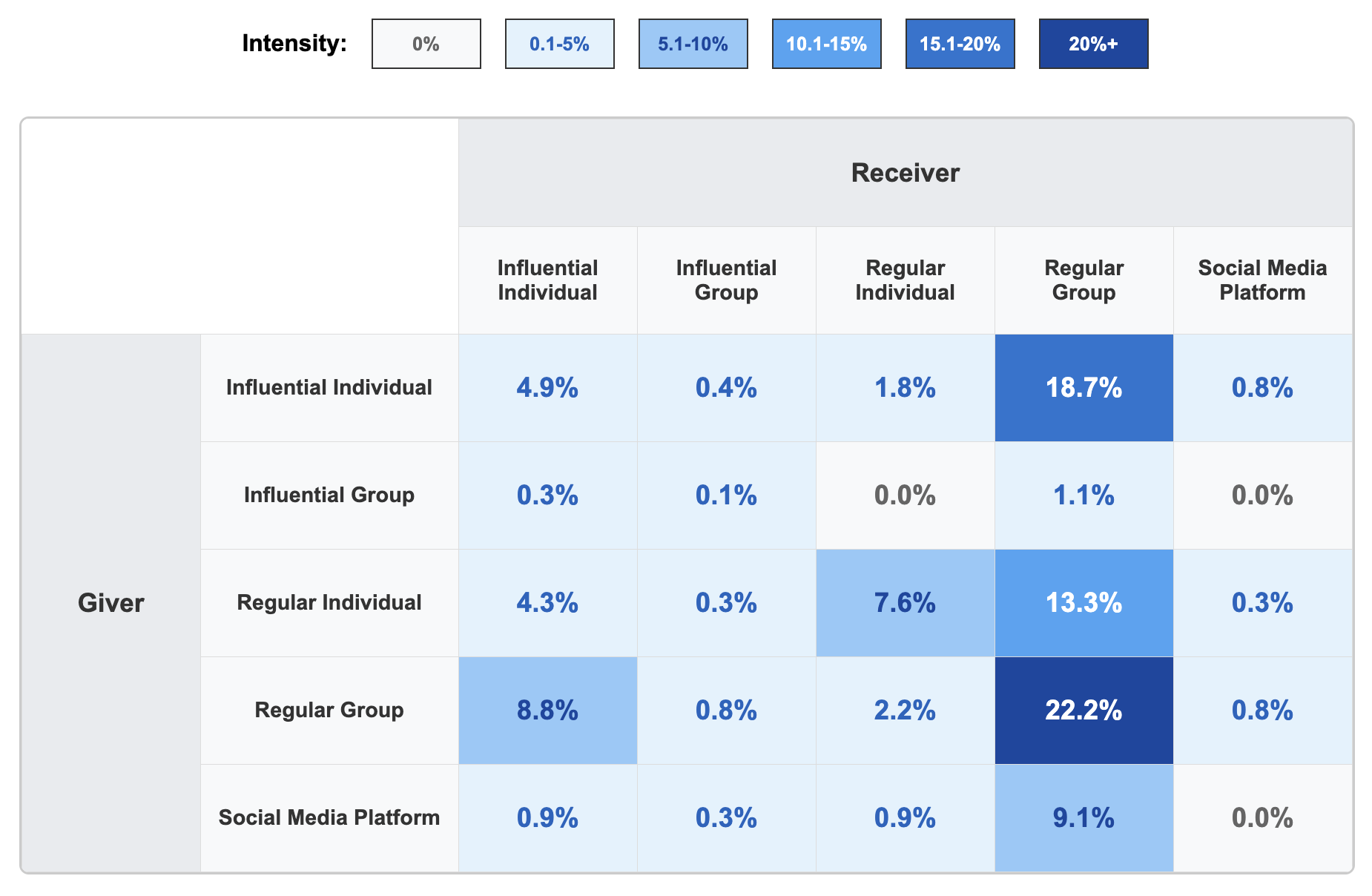}
    \caption{Stakeholder Relationship Matrix (Percentage)}
    \Description{The figure shows the stakeholder relationship matrix according to the percentage among the five possible stakeholders of Influential Individual, Influential Group, Regular Individual, Regular Group, and Social Media Platform. The right column is denoted as the Giver, while the top row is denoted as the Receiver. Regular Group to Regular Group is the most frequent at 22.2 percent, and Influential Individual to Regular Group is second at 18.7 percent.}
    \label{fig:stakeholder-analysis-matrix}
\end{figure}

Though online problematic behaviors may be complex and multi-directional, our analysis found that the vast majority of articles indicated a perpetrator and a target, which we refer to as \stakeholder{Givers} and \stakeholder{Receivers}.
As shown in Figure~\ref{fig:stakeholder-analysis-count} and Figure~\ref{fig:stakeholder-analysis-matrix}, the most prevalent stakeholder type across both \stakeholder{Giver} and \stakeholder{Receiver} roles is the \stakeholder{Regular\_Group}~---~a group of users without large individual influence. Specifically, \stakeholder{Regular\_Group} stakeholders appear in 388 \stakeholder{Giver} cases and 716 \stakeholder{Receiver} cases, far exceeding any other stakeholder category. This indicates that social media crises frequently originate from and impact groups of users who do not hold large influence but participate actively in collective behaviors.

Accordingly, the most common stakeholder interaction pattern is \stakeholder{Regular\_Group} → \stakeholder{Regular\_Group}, accounting for 247 cases (22.2\%). This pattern reveals a key structural feature of \smc{social media crises}: many incidents are community-driven and community-directed, emerging from interactions among collectives rather than influential individuals. 
Other frequent patterns include \stakeholder{Influential\_Individual} → \stakeholder{Regular\_Group} (18.7\%), \stakeholder{Regular\_Individual} → \stakeholder{Regular\_Group} (13.3\%), and \stakeholder{Social Media Platform} → \stakeholder{Regular\_Group} (9.1\%), further illustrating the diverse directions of influence and harm in these crises while including \stakeholder{Regular\_Group} prevalently.

\subsubsection{Articles on Each Online Problematic Behavior}
\begin{figure}
    \centering
    \includegraphics[width=0.6\linewidth]{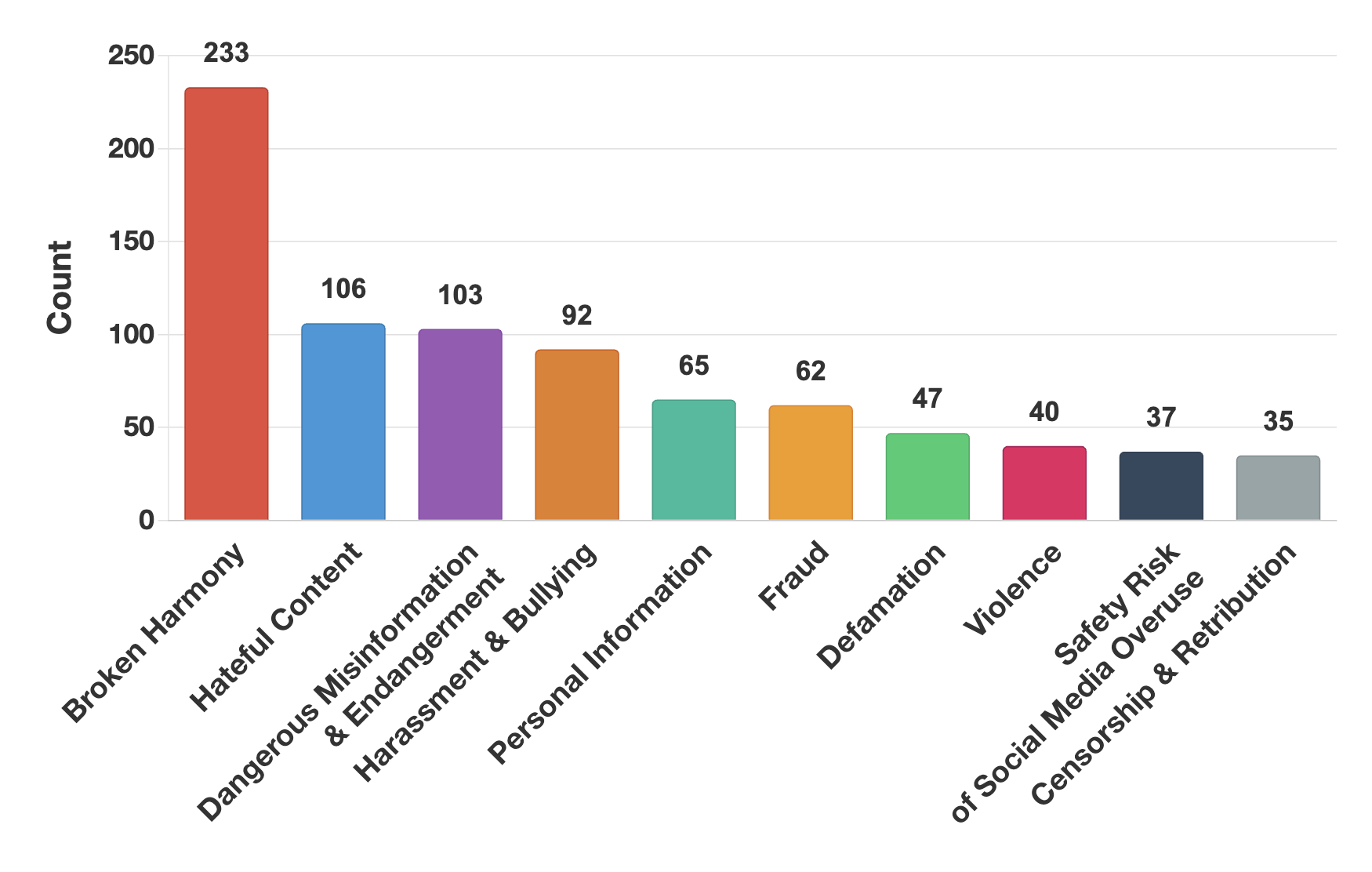}
    \caption{Top 10 Online Problematic Behaviors in CrisisNews}
    \Description{The figure shows a vertical bar graph of the top 10 online problematic behaviors in CrisisNews dataset, with the Broken Harmony at the highest at 233 counts, followed by Hateful Conduct at 106 and Dangerous Misinformation and Endangerment at 103.}
    \label{fig:top-opb}
\end{figure}
Figure~\ref{fig:top-opb} presents the distribution of annotated events across different predefined subcategories of online problematic behaviors. \category{\textbf{User Safety~--~Broken Harmony}} is the most frequently observed subcategory, with 233 instances in the dataset, underscoring the central role of rapid content amplification in contemporary digital harms. This is followed by the subcategories \category{Offensive \& Objectionable Content~---~Hateful Content} and \category{User Safety~--~Dangerous Misinformation \& Endangerment}, each comprising 105 and 93 events, respectively, reflecting the prevalence of harms in the \category{User Safety}. Other common subcategories include Hateful Content, Personal Information, Fraud, and Defamation, all of which show more than 60 events.

\subsection{Platform Involvement in Social Media Crisis}
Among the stakeholder categories identified in our dataset, we focus on cases involving social media platforms as direct stakeholders for several critical reasons. First, platform involvement in crisis events represents a fundamental shift from traditional conceptualizations of platforms as neutral infrastructures to active participants in harm generation and mitigation. Second, when platforms themselves become stakeholders, the scale and impact of resulting crises often extend far beyond typical user-to-user interactions, affecting millions of users simultaneously and triggering regulatory responses. Third, understanding platform behavior as a stakeholder is essential for informing platform governance, policy development, and trust and safety system design.

Our investigation into how social media platforms function as stakeholders within 145 \smc{social media crisis} events (11.1\%) reveals two distinct patterns of engagement (1) cases where platforms serve as givers of harm through mechanisms of systematic violations of user privacy and personal information, censorship and content restriction practices, and algorithmic content moderation failures, and (2) cases where platforms are receivers of harm, involving technically sophisticated and geopolitically significant incidents such as attacks that target critical digital infrastructure, potentially catastrophic in scope and impact. In this section, we focus on the most prevalent cases~---~cases where social media platforms violate user privacy and cases where platforms receive technically sophisticated attacks.

\subsubsection{Social Media Platforms Breach User Trust} The directional analysis reveals a striking asymmetry in how social media platforms function as stakeholders in crisis events. Of the 145 cases involving platforms as stakeholders, 124 cases (85.5\%) positioned the platform as \stakeholder{Givers}, while only 21 cases (14.5\%) involved platforms as \stakeholder{Receivers}. For social media platforms functioning as givers of harm, the most prevalent subcategory (\category{User Safety~--~Personal Information Violations}) involves platforms' handling of user personal information, accounting for 33 cases. These violations demonstrate how platform business models and technical architectures can breach user privacy expectations. Facebook dominates this category with documented incidents spanning from the 2011 revelation that ``Like'' buttons tracked users across the web~\cite{facebook_like} to the massive 2018 Cambridge Analytica data harvesting scandal that affected 87 million users globally~\cite{cambridge}. The 2013 case titled ``Facebook desvela datos de su ayuda al espionaje (Facebook unveils details of its role in espionage)''~\cite{facebook_espionage_2013} revealed the platform's cooperation with NSA surveillance programs, while the 2018 article ``Are you ready? This is all the data Facebook and Google have on you''~\cite{facebook_google_data_2018} provided concrete documentation of the extensive personal data collection practices employed by these platforms.

The escalating regulatory response to these violations is evident in our temporal analysis. Early privacy breaches resulted in public criticism and minor policy adjustments, but recent incidents have generated unprecedented financial penalties. The 2023 case ``€648m fine (largest in Facebook's history)''~\cite{facebook_gdpr_fine_2023} under GDPR regulations demonstrates how privacy violations now carry substantial material consequences for platform operations, representing a fundamental shift in the cost-benefit calculation of aggressive data collection practices.

\subsubsection{Social Media Platforms Are Under Constant Threats}
While platforms rarely function as receivers of harm, one important category of social media platforms as a \stakeholder{Receiver} of harm is \category{Scaled Abuse}. Four documented hacking attempts reveal an escalating arms race between attackers and platform defenses. The 2009 case ``Twitter shut as hackers bombard it with spam''~\cite{twitter_spam_attack_2009} represents early-era volume-based attacks that overwhelmed platform capacity through brute force methods. By 2013, platforms faced more sophisticated threats, as evidenced by ``Facebook Says Hackers Breached Its Computers''~\cite{facebook_breach_2013}, which documented Advanced Persistent Threat (APT) operations involving sustained, coordinated attacks on platform infrastructure.

Recent incidents demonstrate how attack vectors have evolved toward insider threats and privilege escalation. The 2023 case ```GodMode' access is still a problem at Twitter''~\cite{twitter_godmode_2023} reveals how internal administrative privileges can be exploited to compromise platform integrity, suggesting that traditional perimeter security models are insufficient for protecting modern social media infrastructure.

\subsection{User Influence on Social Media Crises}
To further understand the power dynamics within \smc{social media crisis} events, we examined stakeholders based on their level of influence, distinguishing between \stakeholder{Influential} actors and \stakeholder{Regular} actors. This analysis addresses a fundamental question in social media crisis research: who holds the power to initiate and shape crisis events in digital environments?

Our analysis focuses on three key patterns that emerged from our dataset: (1) the dominance of regular users as crisis initiators; (2) the emergence of group-to-group dynamics that reflects the collective nature of digital harm; and (3) the dual role of influential individuals as both perpetrators and victims. These patterns collectively demonstrate that \smc{social media crises} require attention to the complex power dynamics and stakeholder relationships that shape how crises emerge, escalate, and propagate across digital platforms.



\begin{figure}[h]
    \centering
    \includegraphics[width=\textwidth]{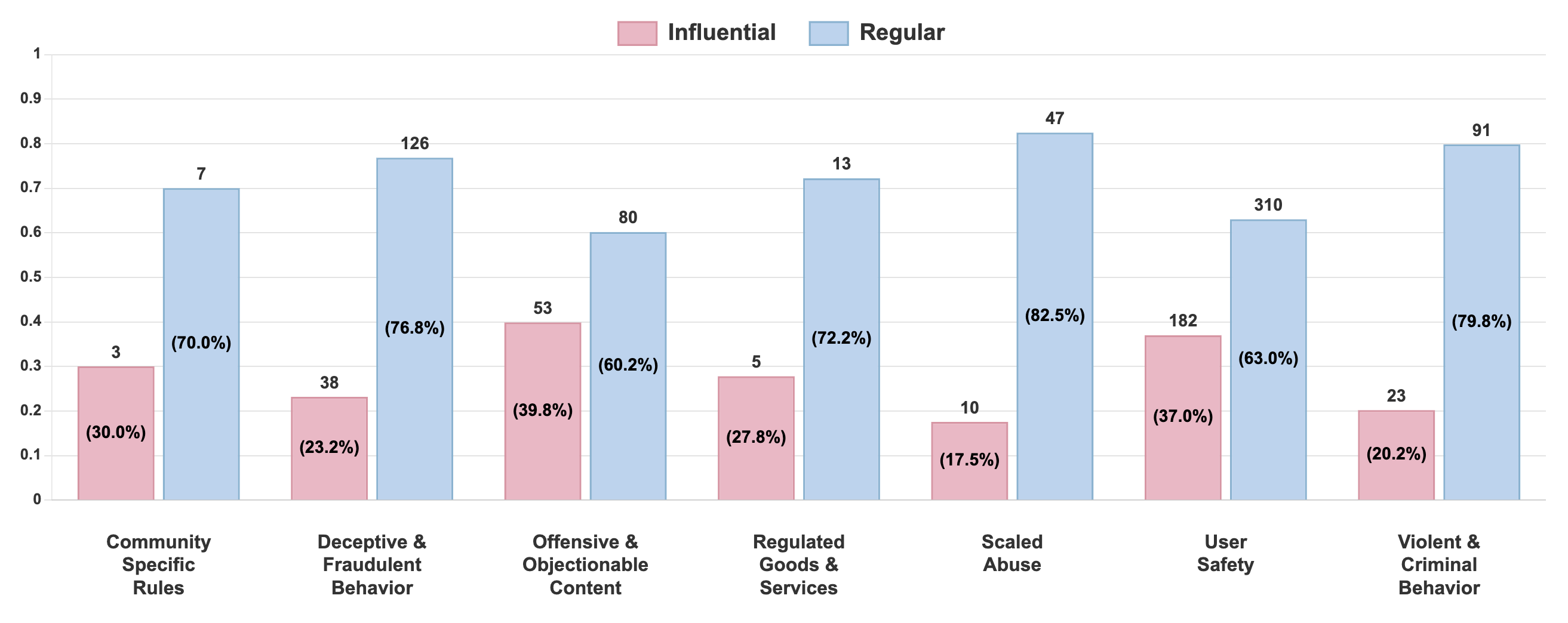}
    \caption{Influential vs. Regular Givers by Online Problematic Behavior Category}
    \Description{The figure shows a vertical bar graph comparing the influential and regular giver counts and percentages by online problematic behavior category. On the left are the influential givers in red bars, and on the right are the regular givers in blue bars for each category. Overall, the regular givers are typically around 60 to 80 percent, while influential givers are between 15 to 40 percent.}
    \label{fig:Giver-Impact}
\end{figure}

\begin{figure}
    \centering
    \includegraphics[width=\linewidth]{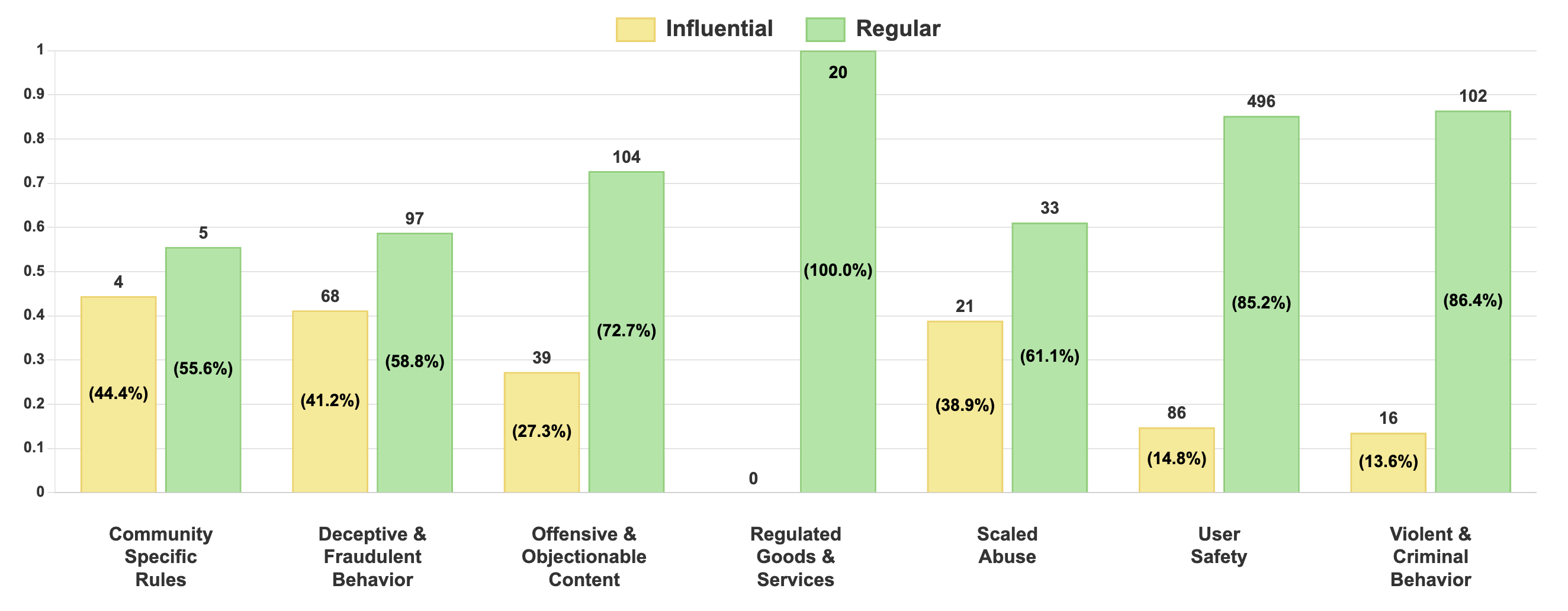}
    \caption{Influential vs. Regular Receivers by Online Problematic Behavior Category}
    \Description{The figure shows a vertical bar graph comparing the influential and regular receiver counts and percentages by online problematic behavior category. On the left are the influential receivers in yellow bars, and on the right are the regular receivers in green bars for each category. Overall, the regular receivers are typically around 50 to 100 percent, while influential givers are between 0 to 40 percent. Notably, for Regulated Goods and Services, all recievers are regular.}
    \label{fig: Receiver-Impact}
\end{figure}


\subsubsection{Regular Stakeholders Have Most Impact as Givers} The analysis of stakeholder impact reveals a consistent pattern in \stakeholder{Regular} stakeholders across both \stakeholder{Giver} and \stakeholder{Receiver} (Figure~\ref{fig:Giver-Impact} and Figure~\ref{fig: Receiver-Impact}). For \stakeholder{Giver}, \stakeholder{Regular} stakeholders constitute the majority across all online problematic behavior categories, with proportions ranging from 60\% to 83\% (Figure~\ref{fig:Giver-Impact}). 
The dominance of \stakeholder{Regular} users as Givers across all online problematic behavior categories challenges common assumptions about who initiates a \smc{social media crisis}. Rather than being primarily driven by high-profile individuals seeking attention or wielding influence, our data suggests that \smc{social media crisis} events may most frequently originate from everyday users engaging in problematic behaviors. Our dataset reveals that these \stakeholder{Regular} stakeholders can generate consequences ranging from individual harm to nationwide policy changes, illustrating the unprecedented decentralization of both crisis generation and resolution mechanisms. Here, decentralization refers to how \stakeholder{Regular} stakeholders now possess the technical capability to create and distribute content that can reach millions of users without editorial oversight or institutional gatekeeping. The viral nature of social media platforms shows that seemingly minor interpersonal conflicts can escalate into significant social phenomena requiring governmental or corporate intervention.

This pattern is most pronounced in \category{Scaled Abuse} category (83\%), where coordinated harmful activities are typically executed by networks of regular accounts rather than prominent figures. The 2018 case ``Your country needs you to fight fake news, UK journalists told''~\cite{last2018} illustrates how regular-appearing accounts, rather than verified influential users, formed the backbone of large-scale manipulation campaigns.
The finding that \stakeholder{Regular} stakeholders predominate even in categories like \category{Offensive and Objectionable Content} (60\%) suggests that the amplification mechanisms of social media platforms can transform ordinary user actions into significant social phenomena. A single offensive post by a regular user can escalate into a widespread crisis when platform algorithms and user sharing behaviors amplify the content beyond its original scope~\cite{collabo23, disinfocscw}.
This pattern reflects a fundamental characteristic of social media ecosystems: the decentralization of influence means that any user can potentially trigger events with far-reaching consequences~\cite{decentral24}, regardless of their initial social standing or follower count.\\

\subsubsection{Group Influence Shifts Social Media Interactions} Among the types of stakeholder relationships that include \stakeholder{Regular} stakeholders, the \stakeholder{Regular\_Group}→\stakeholder{Regular\_Group} relationship demonstrates a fundamental shift in how \smc{social media crises} emerge and propagate. Adding on the point that \stakeholder{Regular} stakeholders could be similarly impactful in social media platforms, we emphasize in this section that the collective actions of multiple \stakeholder{Regular} individuals can create emergent behaviors that exceed the intended impact of any individual participant, also being the most prevalent stakeholder relationship in our dataset.

Our analysis identifies three dominant problematic behavior categories that reveal distinct patterns of how \stakeholder{Regular} stakeholders both create and respond to crises in digital environments in \stakeholder{Regular\_Group}→\stakeholder{Regular\_Group} relationship. The prevalence of \category{User Safety~--~Dangerous Misinformation \& Endangerment} (13.6\%) and \category{User Safety~--~Broken Harmony} (13.2\%) reveals how \stakeholder{Regular} stakeholders can inadvertently or deliberately become vectors for information that leads to tangible harm. Our dataset documents sophisticated coordination between \stakeholder{Regular} stakeholders across multiple platforms to amplify misinformation campaigns. The 2022 case involving ```Troll factory' spreading Russian pro-war lies online''~\cite{russian_troll_factory_2022} reveals how ordinary users can be recruited into organized disinformation operations that span Twitter, Facebook, and other platforms. Similarly, \category{User Safety~--~Broken Harmony} represents a uniquely social media phenomenon where content virality disrupts established social order and individual privacy. The significance lies in how ordinary users now possess unprecedented power to influence public discourse and accountability mechanisms. Notably, we identify the duality of the influence of ~\stakeholder{Regular} stakeholders through contrasting outcomes that demonstrate both documenting misconduct or crime (e.g., ``80-year-old woman [was] thrashed by bahu, video goes viral''~\cite{elderly_woman_abuse_viral} leading to the perpetrator's arrest) and destructive capacity (``Capitals' Evgeny Kuznetsov shown in video next to lines of white powdery substance''~\cite{kuznetsov_substance_video} leading to misinformation) of viral content.

\subsubsection{Influential Individuals Play Both Sides}
Unlike \stakeholder{Regular} stakeholders, \stakeholder{Influential\_Individual} possess the perceived legitimacy and trust that audiences assign to public figures, which fundamentally alters how their actions propagate and impact across social media~\cite{influenceviral24}. \stakeholder{Influential\_Individual} stakeholders include politicians, celebrities, journalists, content creators, and other public figures who possess established authority or significant follower bases on social media platforms. This authority creates asymmetric crisis potential: when ~\stakeholder{Influential} stakeholders engage in problematic behavior, their established credibility and follower networks can amplify harm exponentially~\cite{disinfocscw} beyond what regular users could achieve. Conversely, their public prominence makes them high-value targets for exploitation, impersonation, and coordinated attacks that leverage their visibility for maximum impact~\cite{creator22, ampli23}.

Our analysis reveals that \stakeholder{Influential\_Individual} are involved in approximately 142 cases (30.6\%) of our dataset, making them the second most prevalent stakeholder type after \stakeholder{Regular} stakeholders, which reflects their impact on social media crisis generation and escalation. Among the 142 cases, we observe that \stakeholder{Influential\_Individuals} function both as a problematic actor (\stakeholder{Giver}) in 102 cases (71.8\%), while also serving as the target or victim (\stakeholder{Receiver}) in 40 cases (28.2\%). This distribution reveals that while authority and visibility enable \stakeholder{Influential} stakeholders to shape public discourse, it also makes them vulnerable targets for exploitation and attack. The cases where an \stakeholder{Influential\_Individual} functions as a \stakeholder{Giver} reveal three primary mechanisms through which their authority and reach amplify problematic behaviors into societal crises: weaponizing trust through deceptive practices, exploiting authority to spread offensive content, and breaching information gatekeeping responsibilities. On the flip side, the cases where an \stakeholder{Influential\_Individual} serves as a Receiver reveal systematic targeting patterns that exploit both their public visibility and their accumulated social capital. These attacks represent strategic attempts to hijack influencer credibility for malicious purposes or to damage public figures through coordinated harassment campaigns.

\subsection{Online Problematic Behavior: Broken Harmony}

Among the various types of online problematic behaviors examined in our study, we focus here in greater detail on \category{User Safety~--~Broken Harmony}. This focus is motivated by two factors: first, \category{Broken Harmony} emerged as the most prevalent form of online problematic behavior within \smcdata{}; second, it represents a distinct phenomenon that reveals how social media's viral amplification mechanisms can dramatically escalate the impact of content~---whether initially benign or problematic~---~into sources of widespread social disruption.  We define Broken Harmony as a phenomenon in which the overall cohesion and peaceful state of a social media platform are disrupted. In this context, harmony refers to an environment where users interact and produce content without inflicting harm on others through discriminatory, hateful, or otherwise damaging behaviors. This characteristic, which emphasizes harmony among individuals and within the community, is particularly valued in Asian societies~\cite{asia_01, asia_02} and is less evident in Western approaches to content moderation schemes. 

In this category, we observed events that especially entail the viral spread of controversial content, which triggers widespread public backlash, aggressive discourse, and, in some cases, escalates into tangible consequences such as public apologies, loss of employment or educational opportunities, and organized protest movements (frequently manifesting as hashtag campaigns). From the perspective of social media platforms, the emergence of such aggressive controversy represents a significant risk, as it alters the communicative tone of the space and can foster a persistently negative atmosphere~\cite{pervase}. Given these potential harms, it is critical to systematically examine events that fracture the harmony of social media environments. 

\subsubsection{The harm of exposure to unwanted controversy.}
Certain events are characterized by the rapid, unfiltered spread of socially problematic content on social media, particularly videos of potential violence like public protests or arrests that can inflict real psychological harm on audiences. While news coverage of such events typically emphasizes the content of the incidents themselves, often highlighting illegal or socially deviant behavior, it is essential to recognize that exposure to this material carries the risk of adverse mental health outcomes. Prior research has demonstrated that viewing violent or aggressive content can provoke psychological distress, including symptoms of trauma and post-traumatic stress disorder (PTSD)~\cite{steiger2021psychological}. Accordingly, understanding not only the factual circumstances of these events but also their potential psychological impacts on content consumers is critical for assessing the full scope of harm associated with viral media.

\subsubsection{Viral content can bring positive change in society.}
Meanwhile, there are instances where the disruption of platform harmony through viral controversy ultimately produces constructive societal outcomes. In these cases, although the virality and collective outrage associated with the content may initially pose psychological risks to users, the eventual aftermath contributes to positive change in the perspective of society. One illustrative example involves the circulation of videos documenting illegal or ethically objectionable behavior, which, once widely disseminated, attract the attention of relevant authorities and result in tangible accountability measures such as arrests~\cite{arrest_01, arrest_02, arrest_03} or professional sanctions~\cite{suspend_01, suspend_02}. Moreover, viral controversies have, in some cases, driven broader social reforms, including policy changes, public apologies, and increased awareness of systemic issues such as racial discrimination. Hashtag movements and other forms of digitally coordinated activism frequently emerge in response to such events, underscoring the complex and sometimes beneficial role that viral content can play in shaping public discourse and institutional responses.

%% file: 05_Discussion.tex
\section{Discussion}
\label{sec:Discussion}
This research provides an empirical foundation for understanding how online problematic behaviors evolve into \smc{social media crises} and demonstrates the utility of systematically collecting, categorizing, and analyzing such events. In this section, we highlight important considerations and potential avenues for future research.

\subsection{Reflections on CrisisNews}
The construction of \smcdata{} not only enabled us to aggregate reporting on a broad collection of \smc{social media crises}, but also revealed important insights about the ways such events are represented, documented, and understood. Beyond serving as raw data for analysis, the process of annotation surfaced recurring patterns, gaps, and biases that shape how a \smc{social media crisis} is framed and perceived through journalistic and institutional accounts. These reflections are central to understanding both the strengths and limitations of the dataset, and they highlight methodological considerations relevant to future research on online harms and crisis documentation, which are further discussed in Section~\ref{diss:utilizejournal}.

Our dataset is grounded in news coverage rather than search-based discovery, which enabled the identification of hidden but consequential cases. This perspective allows us to capture incidents that may not achieve viral prominence online but nonetheless generate significant local, institutional, or societal impact, both shown in the overall dataset and Relevant to Crisis category. Our comprehensive review of \smc{social media crises} revealed cases that would likely be missed by keyword-based or trending topic research methods, yet had documented societal impact through news coverage.

Our dataset contains numerous examples of events that generated significant local responses despite limited global visibility. These cases demonstrate how social media can serve as a documentation and accountability mechanism in specific contexts. For instance, the 2021 incident where a Haryana official's ``crack the heads of farmers'' statement~\cite{haryana_farmer_statement_2021} went viral locally led to public backlash and official investigation, demonstrating how citizen documentation can enforce accountability within specific jurisdictions. This news-based validation approach can capture localized, but critical, issues that may be systematically undercounted by high-engagement metrics or global trending topics.

Our annotation process also reveals how \smc{social media crises} involve sophisticated coordination across multiple platforms or subtle manipulation tactics that were well illustrated in news sources compared to single-platform analysis. The 2022 case involving a ```Troll factory' spreading Russian pro-war lies online''~\cite{russian_troll_factory_2022} illustrates how coordinated campaigns can span Twitter, Facebook, and other platforms simultaneously. These events require human analysis to identify the cross-platform coordination patterns over time that automated systems monitoring individual platforms might miss.

\subsection{The Value of Utilizing Journalistic Views for Understanding \smc{Social Media Crises}}
\label{diss:utilizejournal}
The systematic study of crisis events has long been recognized as essential for understanding societal vulnerabilities and developing effective intervention strategies. The field of crisis informatics grew out of the recognition that crises, whether natural disasters, terrorist attacks, or technological failures, often mark decisive turning points~\cite{perry2001facing}. From this perspective, researchers have spent decades exploring how crises are detected, how responses are coordinated, and recovery mechanisms, establishing crisis analysis as a fundamental component of disaster preparedness and social resilience.

However, traditional crisis informatics has primarily focused on offline disasters where social media platforms serve as communication tools during well-defined crisis periods~\cite{DonbasMobile, AustraliaFire, HurricaneHCI, HurricaneHarvey, JapanEarth, MyanmarWar, DistressDisclose, RegretCovid}. This approach, while valuable for understanding crisis response mechanisms, creates systematic blind spots when examining crises that originate and evolve within digital platforms themselves. The challenge lies not merely in detecting these events, but in recognizing why certain patterns of online problematic behavior warrant conceptualization as crises requiring urgent societal intervention.

Our research demonstrates that certain patterns of online problematic behavior exhibit the defining characteristics of crisis events: they pose significant threats to social stability, demand immediate intervention, and create conditions of uncertainty that require coordinated response efforts. Although the application of crisis context to social media events requires careful consideration as it shapes how we understand the urgency, severity, and appropriate responses to online harms, we argue that conceptualizing such events as \smc{social media crises} offers analytical and practical value. It not only advances understanding of online problematic behaviors, but also provides a framework for effective prevention and intervention practices.

Moreover, news sources extend the analytical scope of \smc{social media crisis} research. Unlike platform data, which primarily captures user interactions, journalistic accounts often contextualize events from origin to aftermath. This provides a broader view of how crises unfold and translate into larger consequences, which even includes offline consequences. For example, the 2018 lynchings in Mexico linked to WhatsApp rumors~\cite{mexico_lychining}, Facebook’s removal of seven million coronavirus-related misinformation posts in 2020~\cite{covidmisinfo_21}, the 2021 exposure of a far-right extremist in Washington~\cite{extreme_21}, and the 2023 TikTok ``dragon’s breath'' case in Indonesia~\cite{tiktok_23} all demonstrate how online discourse can escalate into offline harm. By incorporating news coverage alongside platform-based datasets, researchers can enrich their analyses, identifying cases that require additional attention and tracing societal impacts that might otherwise remain less visible.


News reporting also facilitates a temporal lens for studying crisis evolution. While some news articles may convey the snapshot of a specific timeline of an event, collecting articles across the lifespan of an event reveals escalation patterns and strategic adaptations by harmful actors that static content analyses often overlook. In addition, journalistic accounts, by documenting the fundamental interrogatives, can illuminate stakeholder interdependence by showing how crises emerge from interactions among users, algorithms, institutions, and policies, rather than reducing them to isolated user behaviors.

Nevertheless, our findings also underscore the importance of supplementing news-based sources with social media data to capture the full spectrum of \smc{social media crises} and the online problematic behaviors involved. Many reports in our dataset referred to platforms generally as ``social media'', obscuring the role of affordances and cultures specific to each platform. Prior research confirms that news rarely addresses platform design or corporate practices, often distorting public perceptions by assigning blame to individuals while leaving structural causes unexamined~\cite{offthehook}. Furthermore, a single article may frame an incident as minor, while social media data might reveal it as part of a sustained harassment campaign against a vulnerable group. By bringing together platform-level evidence such as content trends, engagement patterns, and community practices with news coverage, researchers can situate crises in both their broader societal narratives and their platform-specific dynamics.

Thus, conceptualizing a \smc{social media crisis} through a crisis framework underscores the urgency, scale, and systemic nature of online harms, and news sources offer invaluable narrative structure and visibility into offline consequences for understanding \smc{social media crises} precisely. Yet, their limitations necessitate complementary data from social media platforms to explore the full scope and exact cause of the event. By combining journalistic accounts with platform-level evidence, researchers can capture both the societal framing and the underlying digital dynamics of crises. This integrative approach advances theoretical understanding of online problematic behaviors, improves detection of emerging threats, and supports the development of governance strategies that balance timely intervention with structural accountability.

\subsection{Implications for Trust and Safety and Social Media Governance}
Our dataset yields important implications for interventions in the trust and safety domain and for advancing more robust frameworks for governing digital environments. First, conceptualizing social media phenomena through the lens of crisis offers a productive theoretical foundation for such efforts. Current content moderation approaches often struggle with prioritization decisions, determining which of thousands of daily policy violations require immediate attention versus routine processing~\cite{prioritize25, prioritize19}. 
Here, the crisis framework's emphasis on acute intervention necessity and substantial social impact will help grasp signals such as rapid amplification, cross-platform spillover that will help distinguish routine policy violations from situations demanding acute intervention. Recognizing and acting on these dynamics is essential for governance, as it enables platforms to allocate resources more strategically, mitigate disproportionate harm, and strengthen resilience against future disruptions. 
Crisis patterns that could be shown through analysis on CrisisNews offer empirical foundations for developing anticipatory models of governance that can distinguish between isolated incidents and emerging crises requiring urgent intervention.

The identification of \category{User Safety~--~Broken Harmony} as our most prevalent subcategory in online problematic behavior reveals another critical insight for platform governance: harm increasingly emerges from amplification patterns and contextual factors rather than from content that violates explicit policies. This category suggests that effective crisis prevention requires monitoring not only content compliance on existing platform rules of violence but also amplification dynamics, relationships of stakeholders related to viral content, and broader contextual forces that can transform innocuous material into potentially harmful phenomena. Particularly, amplification can rapidly escalate the visibility and impact of problematic narratives, overwhelm moderation systems, and disproportionately target vulnerable groups~\cite{ampli23, ampli25}, underscoring its potential in the production of online harm. By integrating amplification-aware metrics into governance frameworks, platforms can better anticipate when seemingly minor incidents are likely to spiral into crises, allowing for earlier, more proportionate, and ultimately more effective interventions before they become crises.

Furthermore, our discovery of problematic behaviors involving large groups of users, such as disinformation campaigns or privacy violations that usually evolve through cross-platform coordination underscores the need for proactive detection systems. While prior research highlights the severity of cross-platform behaviors~\cite{cross20, cross22}, our dataset also revealed instances where large-scale involvement of users produced positive outcomes, such as surfacing ethical concerns or prosecuting perpetrators. These findings suggest that detection should move beyond tracking increases in user activity to differentiating the nature of the spread. Indicators such as the speed of cross-platform diffusion, which groups amplify content, and how actors like fact-checkers or law enforcement respond can inform a more effective moderation strategy. Prior work shows that identical content can spread differently across platforms~\cite{cross23}, underscoring the need for collaboration and clear communication among major platforms for effective intervention~\cite{cross20}. Here, developing more sophisticated detection methods that can identify emerging crisis patterns in real-time, potentially integrating cross-platform monitoring and stakeholder network analysis, could enable proactive intervention strategies for larger scale crises. Ultimately, sharing real-time cross-platform signals would shift the field from reactive content analysis toward proactive, cooperative crisis response that is critical for mitigating harms no single platform can address alone.

The systematic study of social media crises thus offers a paradigmatic advance for online harm research, proposing predictive governance frameworks that can identify and address emerging threats before they escalate into events requiring costly societal intervention. This approach not only enhances platform safety but also contributes to broader social resilience by enabling more effective coordination between digital platforms, regulatory institutions, and civil society organizations in addressing the complex challenges of contemporary digital governance.

%% file: 06_Limitation.tex
\section{Limitations and Future Work}
Our research holds some methodological limitations that highlight important pathways for future investigation. First, our dataset is composed of articles from a limited set of news publishers, with a substantial proportion of sources based in Asia and the United States. We focused primarily (though not exclusively) on English-based articles, 
which introduces the potential for regional bias in English-based news platforms. Consequently, the dataset may not fully capture the global landscape of \smc{social media crises}, especially incidents that unfolded in regions where English-language reporting is less prevalent. Also, the temporal nature of news reporting presents inherent limitations, which became particularly evident during the random sampling conducted for our annotation. Many articles were written during ongoing crises or at the very start of such an event, and thus could not capture long-term consequences or resolutions.

Future research should pursue several complementary directions to address these limitations and extend our contributions. Expanding data collection to incorporate more non-English news sources and alternative documentation methods would provide a more comprehensive and culturally diverse understanding of social media crises. Such methods could include platform transparency reports, user surveys, and ethnographic studies to capture perspectives that traditional news coverage may overlook. 
Moreover, by arranging news articles related to a specific event in chronological order and analyzing them across the crisis life cycle from dection of crisis to a recovery phase~\cite{crisis_phase_Mitroff}, it becomes possible to develop a comprehensive understanding of the event, including its underlying causes and aftermath.

Finally, our framework for understanding \smc{social media crises} could be extended to support predictive modeling and simulation-based research. By combining our taxonomy with computational models of information spread and stakeholder behavior, researchers could test intervention strategies and platform design modifications in controlled environments before deploying them at scale. Such efforts will be essential for building more comprehensive datasets and developing evidence-based approaches to social media safety and crisis prevention.

%% file: 07_Conclusion.tex
\section{Conclusion}
In this work, we introduced CrisisNews, a large-scale dataset that maps two decades of news coverage on \smc{\textit{social media crises}}~---~events of online problematic behavior that originate and escalate within social platforms. By systematically collecting and annotating 93,250 articles, we developed a taxonomy of stakeholder roles, behavior types, and outcomes that reveals how crises evolve across time, platforms, and social contexts. Our analysis underscores how crises are not only the result of isolated harmful actions but also of broader socio-technical dynamics such as virality, amplification, and governance gaps. While our dataset necessarily reflects the biases of journalistic framing, it nonetheless provides a valuable resource for identifying hidden yet consequential cases, enabling comparative analyses across crises, and informing platform governance. We hope that this work inspires future research that expands beyond content-level moderation toward proactive, systemic approaches for crisis prevention, thereby contributing to safer and more trustworthy online environments.

%% file: 08_Appendix.tex
\clearpage
\appendix
\section{List of News Publishers}
\label{app:newspublisher}
\begin{table}[ht]
\label{table:publishers}
\caption{The list of news publishers searched.}
\Description{The table shows the list of news publishers searched, divided by four categories of US Major Dailies, Asia News Publishers, European News Publishers, and Technology Magazines. The four categories are further divided into each country, which includes, in order, the United States, China, Singapore, India, South Korea, Japan, Thailand, Asia-Wide, the UK, Ireland, France, Germany, Spain, Europe-Wide, Network World, and Computerworld. On the far right column is the list of publisher titles for each country.}
\begin{tabular}{l|ll}
\textbf{}                                           & \multicolumn{1}{l|}{\textbf{Publication Country}}   & Publisher Title                    \\ \hline
\multirow{5}{*}{\textbf{U.S. Major Dailies}}        & \multicolumn{1}{l|}{\multirow{5}{*}{United States}} & Chicago Tribune                    \\
                                                    & \multicolumn{1}{l|}{}                               & Los Angeles Times                  \\
                                                    & \multicolumn{1}{l|}{}                               & New York Times                     \\
                                                    & \multicolumn{1}{l|}{}                               & Wall Street Journal                \\
                                                    & \multicolumn{1}{l|}{}                               & The Washington Post                \\ \hline
\multirow{15}{*}{\textbf{Asia News Publishers}}     & \multicolumn{1}{l|}{\multirow{2}{*}{China}}         & Xinhua News Agency - CEIS          \\
                                                    & \multicolumn{1}{l|}{}                               & People's Daily                     \\ \cline{2-3} 
                                                    & \multicolumn{1}{l|}{Singapore}                      & The Straits Times                  \\ \cline{2-3} 
                                                    & \multicolumn{1}{l|}{\multirow{2}{*}{India}}         & The Hindustan Times                \\
                                                    & \multicolumn{1}{l|}{}                               & The Times of India                 \\ \cline{2-3} 
                                                    & \multicolumn{1}{l|}{\multirow{2}{*}{South Korea}}   & The Korea Times                    \\
                                                    & \multicolumn{1}{l|}{}                               & Yonhap News Agency                 \\ \cline{2-3} 
                                                    & \multicolumn{1}{l|}{Japan}                          & The Japan News                     \\ \cline{2-3} 
                                                    & \multicolumn{1}{l|}{\multirow{2}{*}{Thailand}}      & Asia News Monitor                  \\
                                                    & \multicolumn{1}{l|}{}                               & The Nation                         \\ \cline{2-3} 
                                                    & \multicolumn{1}{l|}{\multirow{5}{*}{Asia-Wide}}     & BBC Monitoring Asia Pacific        \\
                                                    & \multicolumn{1}{l|}{}                               & BBC Monitoring Central Asia        \\
                                                    & \multicolumn{1}{l|}{}                               & BBC Monitoring South Asia          \\
                                                    & \multicolumn{1}{l|}{}                               & BBC Monitoring Newsfile            \\
                                                    & \multicolumn{1}{l|}{}                               & BBC Monitoring Media               \\ \hline
\multirow{13}{*}{\textbf{European News Publishers}} & \multicolumn{1}{l|}{\multirow{2}{*}{UK}}            & The Guardian                       \\
                                                    & \multicolumn{1}{l|}{}                               & The Daily Telegraph                \\ \cline{2-3} 
                                                    & \multicolumn{1}{l|}{\multirow{2}{*}{Ireland}}       & Irish Times                        \\
                                                    & \multicolumn{1}{l|}{}                               & Irish Examiner                     \\ \cline{2-3} 
                                                    & \multicolumn{1}{l|}{France}                         & Le Monde                           \\ \cline{2-3} 
                                                    & \multicolumn{1}{l|}{\multirow{3}{*}{Germany}}       & Die Tageszeitung                   \\
                                                    & \multicolumn{1}{l|}{}                               & Die Welt                           \\
                                                    & \multicolumn{1}{l|}{}                               & DPA International (English)        \\ \cline{2-3} 
                                                    & \multicolumn{1}{l|}{Spain}                          & El Pais                            \\ \cline{2-3} 
                                                    & \multicolumn{1}{l|}{\multirow{4}{*}{Europe-Wide}}   & BBC Monitoring European            \\
                                                    & \multicolumn{1}{l|}{}                               & BBC Monitoring Former Soviet Union \\
                                                    & \multicolumn{1}{l|}{}                               & BBC Monitoring Newsfile            \\
                                                    & \multicolumn{1}{l|}{}                               & BBC Monitoring Media               \\ \hline
\multirow{5}{*}{\textbf{Technology Magazine}}       & \multicolumn{1}{l|}{\multirow{3}{*}{United States}} & Bloomberg Businessweek             \\
                                                    & \multicolumn{1}{l|}{}                               & InfoWorld                          \\
                                                    & \multicolumn{1}{l|}{}                               & Wired                              \\ \cline{2-3} 
                                                    & \multicolumn{2}{l}{Network World}                                                        \\ \cline{2-3} 
                                                    & \multicolumn{2}{l}{Computerworld}                                                       
\end{tabular}
\end{table}

\clearpage
\section{List of Social Media Platform Keywords for Initial Selection of News Articles}

\begin{table}[ht]
\caption{List of Keywords on Social Media Platforms for Initial Selection of Dataset}
\Description{The table shows the list of keywords on social media platforms for the initial selection of the dataset by year from 2004 to 2023.}
\label{tab:platform keyword}
\begin{tabular}{l|l}
Year & Keywords                                            \\ \hline
2004 & Facebook, Orkut, Myspace, QQ, Games                 \\
2005 & Piczo, Wink, Myspace, Hi5, World of Warcraft        \\
2006 & Club Penguin, YouTube, StudiVZ, Metacafe, iWiW      \\
2007 & Netlog, SchulerVZ, Niconico, Blingee, Werkenntwen   \\
2008 & MeinVZ, Kaixin001, Tuenti, NK.pl, Odnoklassniki     \\
2009 & Twitter, Kaixin001, Facebook, MeinVZ, Taringa!      \\
2010 & Tumblr, MocoSpace, Chomikuj.pl, Twitter, Facebook   \\
2011 & WhatsApp, Tumblr, Weibo, FC2, Ameba Pigg            \\
2012 & Instagram, WhatsApp, Odnoklassniki, Tumblr, Qeep    \\
2013 & WeChat, ASK, Instagram, WhatsApp, VK                \\
2014 & Facebook, YouTube, Twitter (X), OK, LinkedIn        \\
2015 & Facebook, YouTube, Twitter (X), OK, LinkedIn        \\
2016 & Facebook, YouTube, Twitter (X), OK, LinkedIn        \\
2017 & Facebook, YouTube, Twitter (X), OK, Tumblr          \\
2018 & Facebook, YouTube, Twitter (X), OK, Tumblr          \\
2019 & Facebook, Twitter (X), YouTube, OK, LinkedIn        \\
2020 & Facebook, Twitter (X), YouTube, OK, Instagram       \\
2021 & Facebook, Twitter (X), YouTube, Instagram, LinkedIn \\
2022 & Facebook, Twitter (X), YouTube, LinkedIn, Instagram \\
2023 & Facebook, Twitter (X), YouTube, Instagram, LinkedIn
\end{tabular}
\end{table}

\clearpage
\section{List of Social Media-Related Keywords for Filtering Initial Dataset}

\begin{table}[ht]
\caption{A List of Keywords Used for Dataset Filtering}
\Description{The table shows the list of keywords used for dataset filtering in four columns in alphabetical order.}
\label{tab:keyword}
\begin{tabular}{llll}
\multicolumn{4}{c}{Keywords}                                                                                                                                       \\ \hline
\multicolumn{1}{l|}{Account}                   & \multicolumn{1}{l|}{Geotagging}          & \multicolumn{1}{l|}{Online influencers}       & Social Network service \\
\multicolumn{1}{l|}{Algorithm}                 & \multicolumn{1}{l|}{Hashtags}            & \multicolumn{1}{l|}{Online marketing}         & Stories                \\
\multicolumn{1}{l|}{Chatrooms}                 & \multicolumn{1}{l|}{Instant messaging}   & \multicolumn{1}{l|}{Online polls}             & Streaming              \\
\multicolumn{1}{l|}{Comments}                  & \multicolumn{1}{l|}{Internet}            & \multicolumn{1}{l|}{Online profiles}          & Streaming platforms    \\
\multicolumn{1}{l|}{Content creators}          & \multicolumn{1}{l|}{Internet censorship} & \multicolumn{1}{l|}{Platform}                 & Subscribers            \\
\multicolumn{1}{l|}{Content moderation}        & \multicolumn{1}{l|}{Likes}               & \multicolumn{1}{l|}{Podcast}                  & Tagging                \\
\multicolumn{1}{l|}{Content sharing platforms} & \multicolumn{1}{l|}{Livestream}          & \multicolumn{1}{l|}{Posts}                    & Trending               \\
\multicolumn{1}{l|}{Content strategy}          & \multicolumn{1}{l|}{Livestreaming}       & \multicolumn{1}{l|}{Privacy breach}           & Trolls                 \\
\multicolumn{1}{l|}{Cyberbullying}             & \multicolumn{1}{l|}{Media}               & \multicolumn{1}{l|}{Privacy settings}         & User behavior          \\
\multicolumn{1}{l|}{Data privacy}              & \multicolumn{1}{l|}{Media consumption}   & \multicolumn{1}{l|}{Profile}                  & User-generated content \\
\multicolumn{1}{l|}{Digital footprints}        & \multicolumn{1}{l|}{Memes}               & \multicolumn{1}{l|}{Profile picture}          & Verified accounts      \\
\multicolumn{1}{l|}{Digital marketing}         & \multicolumn{1}{l|}{Mobile}              & \multicolumn{1}{l|}{Reels}                    & Video                  \\
\multicolumn{1}{l|}{Direct message}            & \multicolumn{1}{l|}{Mobile apps}         & \multicolumn{1}{l|}{Selfies}                  & Video sharing          \\
\multicolumn{1}{l|}{DM}                        & \multicolumn{1}{l|}{Multimedia content}  & \multicolumn{1}{l|}{Shares}                   & Video upload           \\
\multicolumn{1}{l|}{Engagement}                & \multicolumn{1}{l|}{Network sharing}     & \multicolumn{1}{l|}{Social media}             & Viral                  \\
\multicolumn{1}{l|}{Engagement rates}          & \multicolumn{1}{l|}{Notifications}       & \multicolumn{1}{l|}{Social media influencers} & Vlogger                \\
\multicolumn{1}{l|}{Fake news}                 & \multicolumn{1}{l|}{Online}              & \multicolumn{1}{l|}{Social media platforms}   & Web                    \\
\multicolumn{1}{l|}{Followers}                 & \multicolumn{1}{l|}{Online communities}  & \multicolumn{1}{l|}{Social Network}           & Website               
\end{tabular}
\end{table}

\section{\textbf{Prompt used for labeling}}
\label{appendix:prompt}

The following is the prompt used for GPT-4o.  

\begin{figure}[ht]
\label{fig:prompt}
\caption{Prompt Used for Labeling Process}
\Description{The figure shows the full text prompt used for GPT-4o labeling process.}
\begin{tcolorbox}[
  colback=black!0!white, colframe=black!20!white, colbacktitle=black!10!white, coltitle=blue!20!black ]

\footnotesize{
\#\#\#Task: You are a critical thinker capable of professional labeling of datasets on news.
Identify three categories for the given row.

Title: \{\textit{(title of the row given as context)}\}, 

\#\#\#<Label 1>: yes or no 

For label1: Read the title and think carefully if the title conveys that the event occurred due to the presence of social media. Here, we will think of the scope of social media as web-based applications and interactive communities that facilitate the creation, discussion, modification, and exchange of user-generated content, thus not only including SNS communities such as X or Instagram but also including messenger applications such as WhatsApp or Telegram.
Label as
- yes if the title conveys the presence of a social media community causing such an event. If the title conveys the presence of online space while not including a specific social media community, but contains a potential that the event included a social community online, also label as ‘yes’.
- no if the title conveys the event did not occur because of social media.\\

\#\#\#<Label 2>: yes or no

For label2: Read the title and think carefully if the title involves online problematic behavior.
Label as
- yes if the title conveys the presence of online problematic behavior.
- no if the title does not convey any online problematic behavior.

}
\end{tcolorbox}           
\end{figure}

\clearpage
\section{List of Full Subcategories in Annotation}
\label{fullsubcat}
\subsection{Online Problematic Behavior}
\begin{table}[H]
\caption{A List of Subcategories Used for Annotation}
\Description{The table shows the full list of categories and their subcategories of abuse types used for annotation. The categories are Violent and Criminal Behavior, Regulated Goods and Services, Offensive and Objectionable Content, User Safety, Scaled Abuse, Deceptive and Fraudulent Behavior, and Community-Specific Rules. The subcategories are on the right column.}
\label{tab:annopb}
\begin{tabular}{l|l}
\textbf{Categories of the Abuse Types}      & \textbf{Subcategory}                             \\ \hline
\multirow{5}{*}{Violent \& Criminal Behavior}       & Child Abuse \& Nudity                    \\
                                                    & Sexual Exploitation                      \\
                                                    & Dangerous Organizations                  \\
                                                    & Violence                                 \\
                                                    & Illegal Behavior                         \\ \hline
\multirow{3}{*}{Regulated Goods \& Services}        & Regulated Goods                          \\
                                                    & Regulated Services                       \\
                                                    & Commercial Sexual Activity               \\ \hline
\multirow{3}{*}{Offensive \& Objectionable Content} & Graphic \& Violent Content               \\
                                                    & Hateful Content                          \\
                                                    & Nudity \& Sexual Activity                \\ \hline
\multirow{7}{*}{User Safety}                        & Suicide \& Self-harm                     \\
                                                    & Dangerous Misinformation \& Endangerment \\
                                                    & Personal Information                     \\
                                                    & Broken Harmony                           \\
                                                    & Safety Risk of Social Media Overuse      \\
                                                    & Harassment \& Bullying                   \\
                                                    & Censorship \& Retribution                \\ \hline
\multirow{4}{*}{Scaled Abuse}                       & Hacking                                  \\
                                                    & Malware                                  \\
                                                    & Inauthentic Behavior                     \\
                                                    & Spam                                     \\ \hline
\multirow{4}{*}{Deceptive \& Fraudulent Behavior}   & Fraud                                    \\
                                                    & Intellectual Property                    \\
                                                    & Impersonation                            \\
                                                    & Defamation                               \\ \hline
\multirow{2}{*}{Community-Specific Rules}           & Content Limitation                       \\
                                                    & Format                                  
\end{tabular}
\end{table}

\clearpage
\subsection{Aftermath}
\begin{table}[h]
\caption{A List of Subcategories Used for Aftermath Annotation}
\label{tab:aftermath}
\begin{tabular}{ll}
\multicolumn{2}{c}{\textbf{Aftermath}}                                                     \\ \hline
\multicolumn{1}{l|}{Address Broken Harmony}         & Misinformation Led to Offline Action \\
\multicolumn{1}{l|}{Address Misinformation}         & Negative Impact to Public Property   \\
\multicolumn{1}{l|}{Banned from Social Media}       & Physical Harm                        \\
\multicolumn{1}{l|}{Buy Social Media Content}       & Platform Ban                         \\
\multicolumn{1}{l|}{Commercial Pullout}             & Platform Reputation Damage           \\
\multicolumn{1}{l|}{Content Removal}                & Protective Measures                  \\
\multicolumn{1}{l|}{Create Protective Organization} & Public Apology                       \\
\multicolumn{1}{l|}{Death}                          & Public Backlash                      \\
\multicolumn{1}{l|}{Enforcement}                    & Public Criticism                     \\
\multicolumn{1}{l|}{Real-life Fraud}                & Public Message from/to Social Media  \\
\multicolumn{1}{l|}{Government Initiative}          & Public Protest                       \\
\multicolumn{1}{l|}{Leave Social Media}             & Regulatory Inquiry                   \\
\multicolumn{1}{l|}{Legal Action}                   & Social Media Policy Change           \\
\multicolumn{1}{l|}{Legislation}                    & Violence / Violent Behavior          \\
\multicolumn{1}{l|}{Manipulated Behavior}           & Whistleblow                          \\
\multicolumn{1}{l|}{Mental Health Issues}           &                    \Description{The table shows the list of subcategories used for aftermath annotation in alphabetical order.}                 
\end{tabular}
\end{table}

\clearpage
\subsection{Platform}

\begin{table}[H]
\caption{A List of Subcategories Used for Platform Annotation}
\label{tab:platform}
\begin{tabular}{lll}
\multicolumn{3}{c}{\textbf{Platform}}                                                                       \\ \hline
\multicolumn{1}{l|}{4chan}            & \multicolumn{1}{l|}{Get Transcript}         & RealSelf              \\
\multicolumn{1}{l|}{8chan}            & \multicolumn{1}{l|}{Google}                 & Reddit                \\
\multicolumn{1}{l|}{Activism Website} & \multicolumn{1}{l|}{Google Chat}            & Rutube                \\
\multicolumn{1}{l|}{Airbnb}           & \multicolumn{1}{l|}{Grindr}                 & Salon24               \\
\multicolumn{1}{l|}{Amazon}           & \multicolumn{1}{l|}{Guardian}               & Shopee                \\
\multicolumn{1}{l|}{Apple}            & \multicolumn{1}{l|}{Hotmail}                & SmartTV               \\
\multicolumn{1}{l|}{Ashley Madison}   & \multicolumn{1}{l|}{Iconfactory}            & Snapchat              \\
\multicolumn{1}{l|}{Ask.fm}           & \multicolumn{1}{l|}{Instagram}              & Social Media          \\
\multicolumn{1}{l|}{Badoo}            & \multicolumn{1}{l|}{IsAnyoneUp}             & Social Trade Dot Bizz \\
\multicolumn{1}{l|}{Bebo}             & \multicolumn{1}{l|}{Ketto}                  & Spokeo                \\
\multicolumn{1}{l|}{Bilibili}         & \multicolumn{1}{l|}{Koo}                    & Spotify               \\
\multicolumn{1}{l|}{Blog}             & \multicolumn{1}{l|}{LinkedIn}               & Telegram              \\
\multicolumn{1}{l|}{boAt}             & \multicolumn{1}{l|}{Merriam-Webster}        & TikTok                \\
\multicolumn{1}{l|}{Bukalapak}        & \multicolumn{1}{l|}{Meta}                   & Tokopedia             \\
\multicolumn{1}{l|}{Carousell}        & \multicolumn{1}{l|}{Microsoft}              & Tumblr                \\
\multicolumn{1}{l|}{CBOT}             & \multicolumn{1}{l|}{Mobile Application}     & Tuenti                \\
\multicolumn{1}{l|}{Cobrapost}        & \multicolumn{1}{l|}{Mugshots.com}           & Turkdunya             \\
\multicolumn{1}{l|}{Craigslist}       & \multicolumn{1}{l|}{Muslim Massacre}        & Twitch                \\
\multicolumn{1}{l|}{CSDN}             & \multicolumn{1}{l|}{MySpace}                & Twitter               \\
\multicolumn{1}{l|}{Daily Mail}       & \multicolumn{1}{l|}{News}                   & Viber                 \\
\multicolumn{1}{l|}{Dating App}       & \multicolumn{1}{l|}{OLX}                    & Vkontakte             \\
\multicolumn{1}{l|}{Delphi}           & \multicolumn{1}{l|}{Online}                 & Web Diary             \\
\multicolumn{1}{l|}{Digg}             & \multicolumn{1}{l|}{Online Chat}            & Webtoon               \\
\multicolumn{1}{l|}{Discord}          & \multicolumn{1}{l|}{Online Dating Website}  & WeChat                \\
\multicolumn{1}{l|}{E-Commerce}       & \multicolumn{1}{l|}{Online News Website}    & Weibo                 \\
\multicolumn{1}{l|}{eBay}             & \multicolumn{1}{l|}{Online Resale Platform} & WhatsApp              \\
\multicolumn{1}{l|}{Email}            & \multicolumn{1}{l|}{Orkut}                  & Wikipedia             \\
\multicolumn{1}{l|}{Facebook}         & \multicolumn{1}{l|}{Parler}                 & Xbox Live             \\
\multicolumn{1}{l|}{Fox News}         & \multicolumn{1}{l|}{Pinterest}              & Yahoo                 \\
\multicolumn{1}{l|}{Game}             & \multicolumn{1}{l|}{PUBG}                   & Yandex                \\
\multicolumn{1}{l|}{Gawker}           & \multicolumn{1}{l|}{QQ}                     & Youtube               \\
\multicolumn{1}{l|}{Game}             & \multicolumn{1}{l|}{RateMyTeachers.com}     & Zoom                 
\Description{The table shows the list of subcategories used for platform annotation in alphabetical order.}
\end{tabular}
\end{table}